\documentclass[12pt,preprint]{aastex}

\newcommand{\nhico}{$\frac{N_{H_2}}{I_{CO}}$}

\newcommand{\CO}{$^{12}$CO}
\newcommand{\COe}{$^{12}$CO $1-0$ }

\newcommand{\vol}[1]{1}
\newcommand{\kms}{km\,s$^{-1}$}

\newcommand{\solm}{M$_{\odot}$}

\slugcomment{}

\shorttitle{The Double Barred Galaxy NGC~4303}
\shortauthors{Schinnerer et al.}

\begin{document}

\title{Towards the Secondary Bar: \\
Gas Morphology and Dynamics in NGC~4303}

\author{Eva Schinnerer}
\affil{Department of Astronomy, MS 105-24, California Institute of Technology,
       Pasadena, CA 91125}
\email{es@astro.caltech.edu}

\author{Witold Maciejewski}
\affil{Osservatorio Astrofisico di Arcetri, Largo E. Fermi 5, 50125 Firenze, 
Italy, and 
Obserwatorium Astronomiczne Uniwersytetu Jagiello{\'n}skiego, Poland}
\email{witold@arcetri.astro.it}

\author{Nick Scoville}
\affil{Department of Astronomy, MS 105-24, California Institute of Technology,
       Pasadena, CA 91125}
\email{nzs@astro.caltech.edu}

\and

\author{Leonidas A. Moustakas}
\affil{Dept. of Astrophysics, Nuclear and Astrophysics Laboratory,
University of Oxford, 1 Keble Road, Oxford OX1 3RH, UK} 
\email{leonidas@astro.ox.ac.uk}

\begin{abstract}
We report on the molecular gas properties in the central kiloparsec of
the almost face-on double barred galaxy NGC~4303 (M~61), using the
\COe~line emission observed with the Owens Valley Radio Observatory (OVRO)
millimeter interferometer. The bulk of the molecular line emission
comes from two straight gas lanes which run north-south along the
leading sides of the large-scale primary bar. Velocity deviations of up
to 90 \kms\ from the mean rotational velocity are associated with
these gas lanes. Inside a radius of $\sim5''$ ($\sim400$\,pc) the
molecular gas forms a spiral pattern which, for the northern arm, can
be traced to the nucleus. 
The high angular resolution of our OVRO data ($2'' \sim$ 150~pc),
together with archival HST data, allows for a comparison with
dynamical models of gas flow in the inner kiloparsec of single- and
double-barred galaxies. We find that the observed global properties of
the molecular gas are in agreement with models for the gas flow in a
strong, large-scale bar, and the two-arm spiral structure seen in \CO\
in the inner kiloparsec can already be explained by a density wave
initiated by the potential of that bar. Only a weak correlation
between the molecular gas distribution and the extinction seen in the
HST $V-H$ map is found in the inner 400 pc of NGC~4303: The innermost
part of one arm of the nuclear \CO\ spiral correlates with a weak dust
filament in the color map, while the overall dust distribution follows
a ring or single-arm spiral pattern well correlated with the UV
continuum. This complicated nuclear geometry of the stellar and
gaseous components allows for two scenarios: (A) A self-gravitating
$m=1$ mode is present forming the spiral structure seen in the UV
continuum. In this case the gas kinematics would be unaffected by the
small ($\sim 4''$) inner bar. (B) The UV continuum traces a complete
ring which is heavily extincted north of the nucleus. Such a ring
forms in hydrodynamic models of double bars, but the models cannot
account for the UV emission observed on the leading side of the inner
bar. Comparison with other starburst ring galaxies where the
molecular gas emission and the star forming clusters form a ring or
tightly wound spiral structure suggests that the starburst ring in
NGC~4303 is in an early stage of formation.
\end{abstract}


\section{INTRODUCTION}\label{sec:intro}

Recent studies show that about 75\% of all spiral galaxies have bars,
and out of those, some 40\% have at least one inner (second) bar (see
e.g. Laine et al. 2002, Erwin \& Sparke 2002). The inner bar appears
randomly oriented with respect to the outer, main, bar (e.g. Buta \&
Crocker 1993), indicating that the two bars may rotate independently.
Such systems of nested bars may funnel galactic gas towards an active
galactic nucleus (AGN), as proposed by Shlosman, Frank \& Begelman
(1989). However, Shlosman et al. considered only gaseous inner bars
resulting from the instability in the nuclear disk.  Detection of
nuclear bars in the near infrared (e.g. Jungwiert, Combes
\& Axon 1997) indicates that their predominant component may be stellar. 
Friedli \& Martinet (1993) investigated the stability of stellar
double bars, and found that such systems can survive for a few
rotations after the two bars decouple. Recently, Rautiainen \& Salo
(1999) showed that various modes, such as spiral arms and multiple
bars, can coexist for several Giga-years. Maciejewski \& Sparke (2000)
found orbital families that can support double bars, and that are
related to the $x_1$ and $x_2$ orbits in single barred galaxies. They
predicted that gas flows in dynamically possible stellar nested bars
lack straight shocks in the inner bar, and do not involve strong gas
inflow; they therefore differ significantly from the instability
scenario proposed by Shlosman et al. Given this discrepancy, it is
important to observe the molecular gas in double barred galaxies, so
that its morphology and kinematics can be confronted with the
diverging dynamical scenarios.

Detailed comparison of individual galaxies with high-resolution
dynamical models allows us to explore the mechanisms of gas transport
on large and small scales. Gas condensations in lanes running along
the leading side of a large-scale bar in the models of Athanassoula
(1992) are observed as dust lanes in a number of galaxies.  Recent
dynamical models by Englmaier \& Shlosman (2000) for the gas flow in
the inner kiloparsec of a barred galaxy showed that an induced gas
density wave could form a grand-design structure.  A recent HST survey
of the nuclear regions of nearby Seyfert galaxies shows that spiral
structures (either grand-design or flocculent) are common in the
central kiloparsec (Pogge \& Martini 2002).

Despite the progress in modeling, few observations of the gas flow in
double barred galaxies exist to-date which allow one to carefully test
the validity of the theoretical assumptions. An almost face-on
inclination is favored for comparisons to dynamical models, since
deprojection effects are minimized. Furthermore, the bars should not
lie close to the kinematic major or minor axes in order that
non-circular motion can be more easily studied. For our study, we
selected the nearly face-on galaxy NGC~4303 (M 61).

NGC~4303 is probably associated with the Virgo cluster (Binggeli,
Sandage \& Tammann 1985). A large-scale bar of about 40'' (Laine et
al. 2002) lies inside the outer spiral arms.  In addition to the
large-scale bar, NGC~4303 hosts a second, inner bar of about 4''
length which is surrounded by a circumnuclear starburst ring/spiral of
$6''$ ($\sim$~0.5~kpc) diameter (Colina \& Wada 2000, Perez-Ramirez et
al. 2000, Colina et al. 1997). A high-angular resolution HST UV image
shows several compact star forming regions within this ring. The
starburst ring dominates the integrated UV output of NGC~4303 (Colina
et al. 1997). The nucleus is classified as a LINER/Seyfert~2 type (Ho,
Filippenko \& Sargent 1997). For consistency with the work by Colina
et al. (1997, 1999, 2000), we adopt the distance of M~100, the
brightest spiral in the Virgo cluster (16.1 Mpc, Ferrarese et
al. 1996; $1'' \approx 78$\,pc).

The paper is organized as follows: After a brief description of the
observations in Section~\ref{sec:observations}, the kinematic and
dynamical quantities directly derivable from our mm-interferometric
data are presented in Section~\ref{sec:mmobs}. In
Section~\ref{sec:mmstar}, the molecular gas distribution is compared
to the stellar properties in the inner kiloparsec, and the bar
properties as well as the dynamical resonances are derived. In
Sections \ref{sec:geomdyn} and \ref{subsec:gashst} we interpret the
observed gas morphology and kinematics in the framework of available
dynamical models. Implications and discussion are presented in
Section~\ref{sec:bmods}, followed by the Summary and Conclusions in
Section~\ref{sec:conc}.

\section{OBSERVATIONS}\label{sec:observations}

\subsection{Millimeter Interferometric Observations}

NGC~4303 was observed in its \COe\ line between 1999~December and
2000~March using the Owens Valley Radio Observatory (OVRO) millimeter
interferometer with six 10.4~m telescopes in its E, H and U
configurations. The resulting baselines range from 30~m to 480~m
providing a spatial resolution of $\sim 2.0''$ (150~pc) with natural
weighting (beam size: $2.22'' \times 1.84''$ with $PA$ of
$84.3^o$). Two spectrometer modules were utilized resulting in 120
channels with a spectral resolution of 5.2\,\kms\ per channel. The
quasar 3C273 served as a passband and phase calibrator and was
observed every 20 minutes. The average single sideband temperature was
around 800\,K at the observed line frequency of 114.67\,GHz. The
resulting noise per channel is $\sim 25$\,mJy\,beam$^{-1}$ in the
combined data of the 3 tracks. (The two channels at velocities of
1596\,\kms\ and 1574\,\kms\ have a slightly higher r.m.s. due to the
reduced sensitivity at the edge of one spectrometer module.) The data
were calibrated using the OVRO software package MMA (Scoville et
al. 1993). The final \CO\ cube was mapped and deconvolved using the
CLEAN algorithm within the software package MIRIAD (Sault et
al. 1995). For the deconvolution we used defined CLEAN regions, 10000
iterations, a gain of 0.0075 and a 1$\sigma$ cutoff level of 25
mJy\,beam$^{-1}$, thus allowing for a proper reduction of the beam
effects in the dirty cube. Side lobe artifacts might dilute the line
emission seen in the straight gas lanes running in north-south
direction, as the dirty beam has side lobes at the 35\% level about
$6''$ north and south of the center due to the low declination of the
source ($\sim 4^{\circ}$; Fig.~\ref{fig:COall}). However, since we
detected no line emission above the 5$\sigma$ level in the channel
maps (Fig. \ref{fig:COchan}) on the opposite sites of the gas lanes,
we are confident that all prominent side lobes artifacts have been
removed. Further analysis (moment maps, flux measurements, rotation
curve fitting) was performed using the radio software package GIPSY
(van der Hulst et al. 1992). The data cubes were corrected for the
response of the primary beam, since the outer emission peaks coincide
with the half power beam width (HPBW) of the OVRO dishes. The (biased)
moment maps were made using a clipping level of $3\sigma$ with the
restriction that emission above the clipping level is at least present
in two adjacent channels in the primary beam corrected data cube.

\subsection{HST Archival Data}

Through the STScI HST Archive we obtained high resolution imaging data
of the nuclear region of NGC~4303. The datasets are summarized in
Table~\ref{tab:hstlog}, and include HST/WFPC2 data in the F606W
filter, HST/NICMOS (Camera 2) imaging in the F160W filter (see also
Colina \& Wada 2000), and HST/STIS imaging with the STIS/NUV-MAMA
detector (P.I. Colina). We used standard reduction procedures. The
optical image was resampled to the pixel scale of NIR image. All
images were registered to match the location of the bright central
source with the H band peak emission. The registration of all images
was checked on the dusty regions and the stellar clusters, and showed
good agreement ($\ll 1''$).

\section{THE INTERFEROMETRIC DATA ON THE MOLECULAR GAS}\label{sec:mmobs}

\subsection{Distribution}

\begin{figure*}
\epsscale{}
\plotone{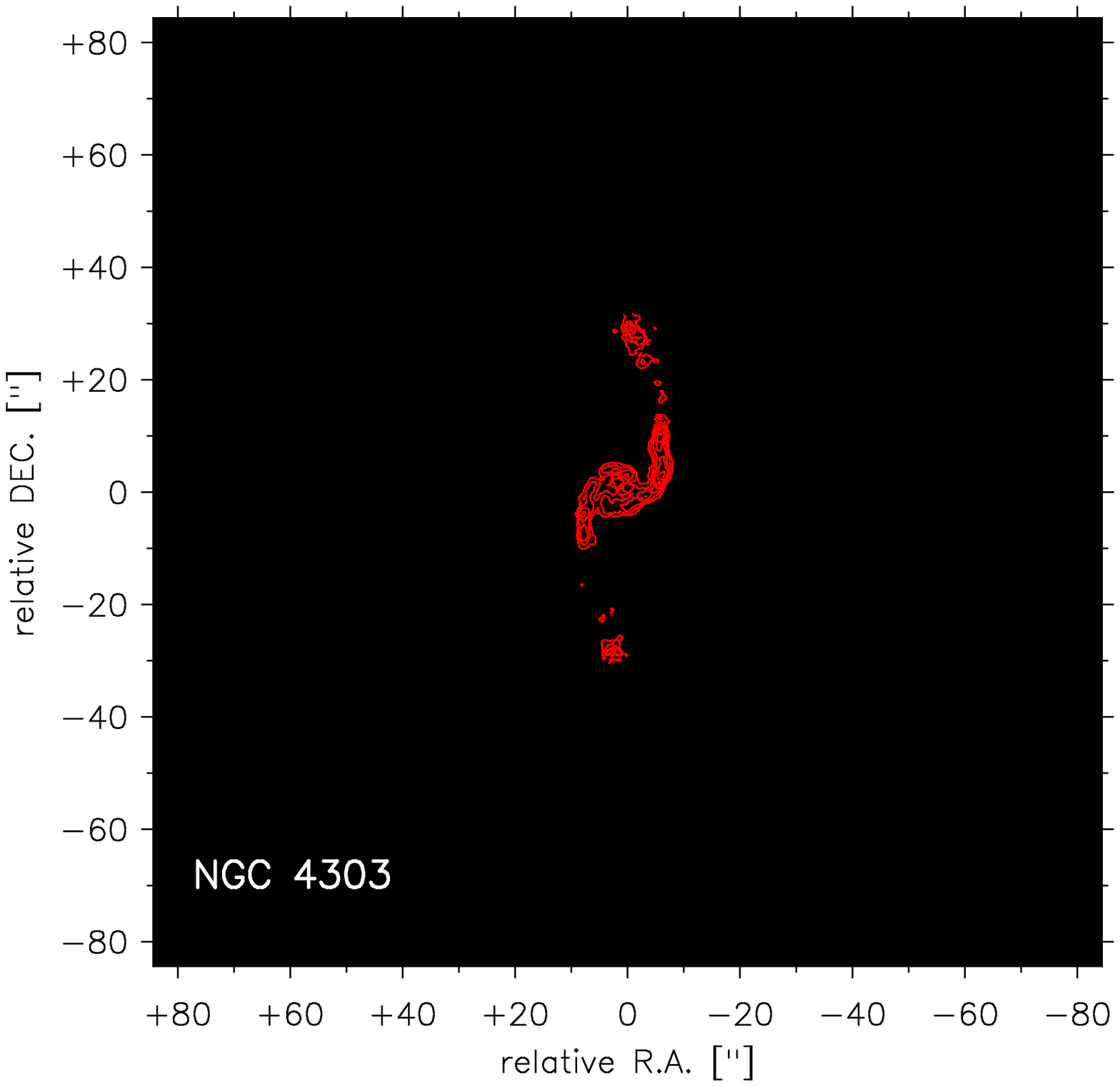}
\caption{ 
The \COe~emission (red contours; primary beam corrected) as observed
with the OVRO mm-interferometer lies inside the prominent spirals seen
in the $i$ band image of Frei et al. (1996). Note, that there is also
molecular emission associated with the spiral arms (see Sheth 2001).
}\label{fig:COiband}
\end{figure*}

The molecular gas emission is concentrated in two straight gas lanes
running north and south of the nucleus (Fig.~\ref{fig:COiband} and
\ref{fig:COall}). These gas lanes show a lateral offset from the
nucleus of about 7$''$ and curl towards the nucleus inside a radius of
$r \sim 5''$. The distribution of the molecular gas emission is not
smooth, showing local contrasts of more than a factor of four within the
gas lanes.  In addition, two peaks of emission about $28''$ (2.0\,kpc)
north and south of the nucleus are present (see
Fig.~\ref{fig:COfet}). Comparison of spectra extracted for these peaks
with the nuclear region rule out the possibility that these peaks
could arise from side lobe artifacts due to the low declination of the
source (Fig.~\ref{fig:CObar}). Those peaks clearly show a velocity shift of
the line centers. The geometry of the molecular gas as observed in the
\COe line emission is reminiscent of the dust lane structures seen in
simulations of the gas response in a barred potential (e.g. Fig.~2 of
Athanassoula 1992; hereafter A92; see Section~\ref{sec:geomdyn}).

\subsection{The kinematics}\label{subsec:kin}

\begin{figure*}
\epsscale{.80}
\plotone{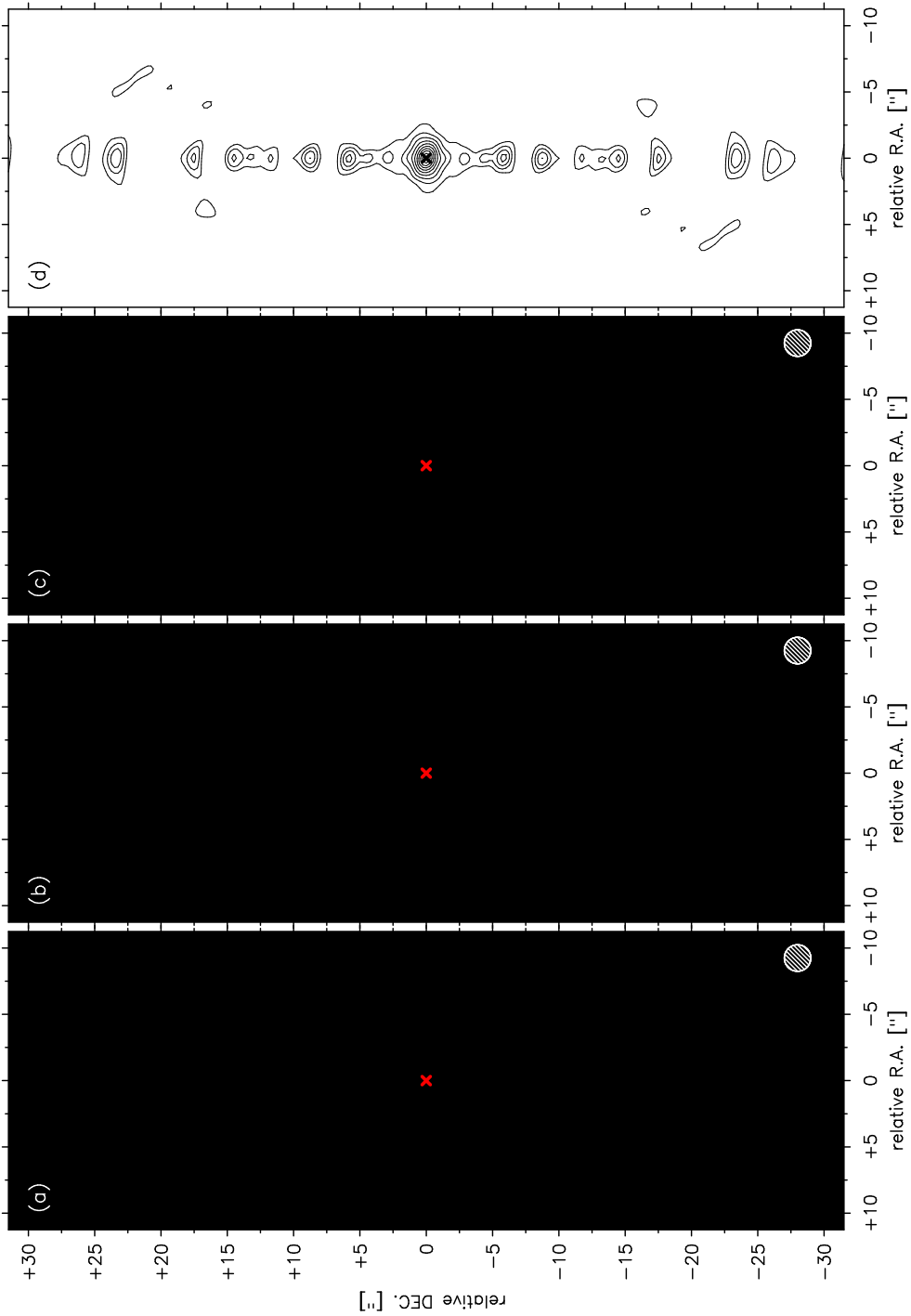}
\caption{ 
Moment maps of NGC~4303 from the OVRO mm-interferometer (primary beam
corrected) \COe ~line data at a spatial resolution of $2.0''$. The
offset coordinates refer to the derived dynamical center (see Table
\ref{ww1}). {\it (a)} Intensity map (0th moment), {\it (b)}
Velocity field (1st moment), and {\it (c)} Dispersion map (2nd
moment). {\it (d)} The dirty been is shown in contours of 10\%, 20\%,
... ,100\%.  }\label{fig:COall}
\end{figure*}

\begin{figure*}
\epsscale{.50}
\plotone{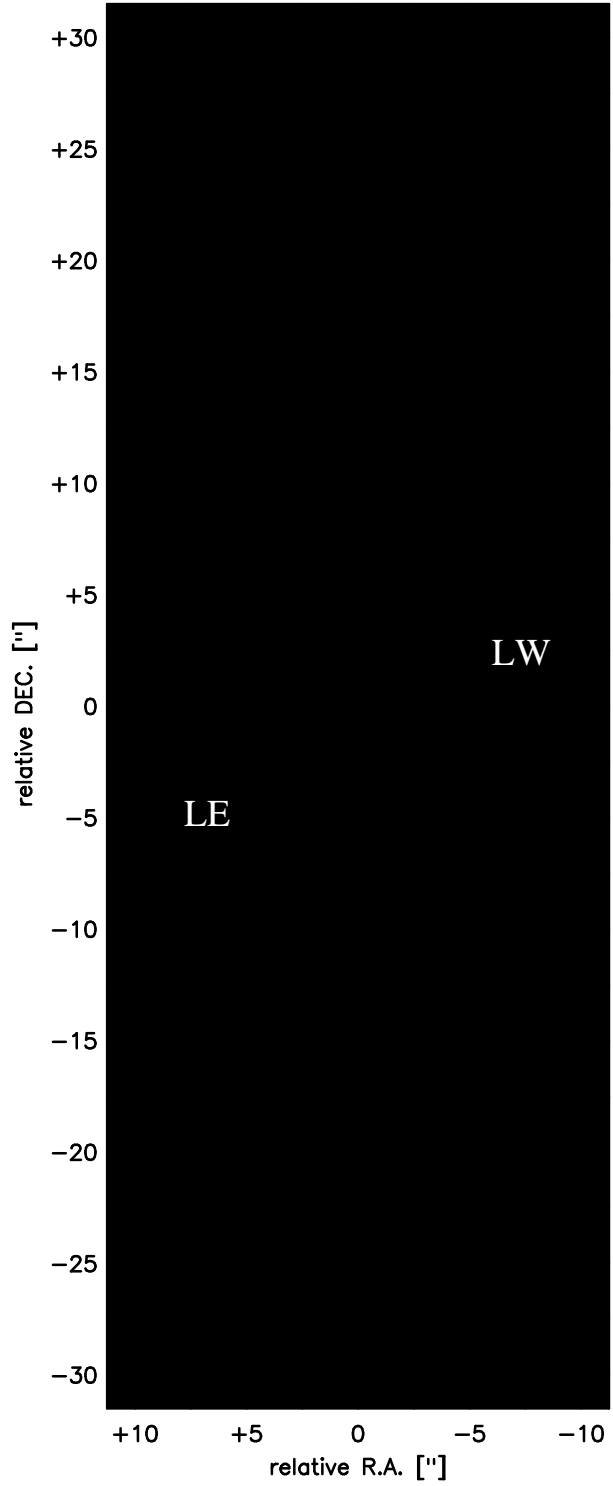}
\caption{
The individual components of the \CO\ line emission which are
discussed in the text are indicated in the 0th moment map. (PN, PS,
LE, LW, and N denote the position of the extracted spectra shown in
Fig. \ref{fig:CObar}.)  }\label{fig:COfet}
\end{figure*}

\begin{figure*}[ht]
\epsscale{1.0}
\plotone{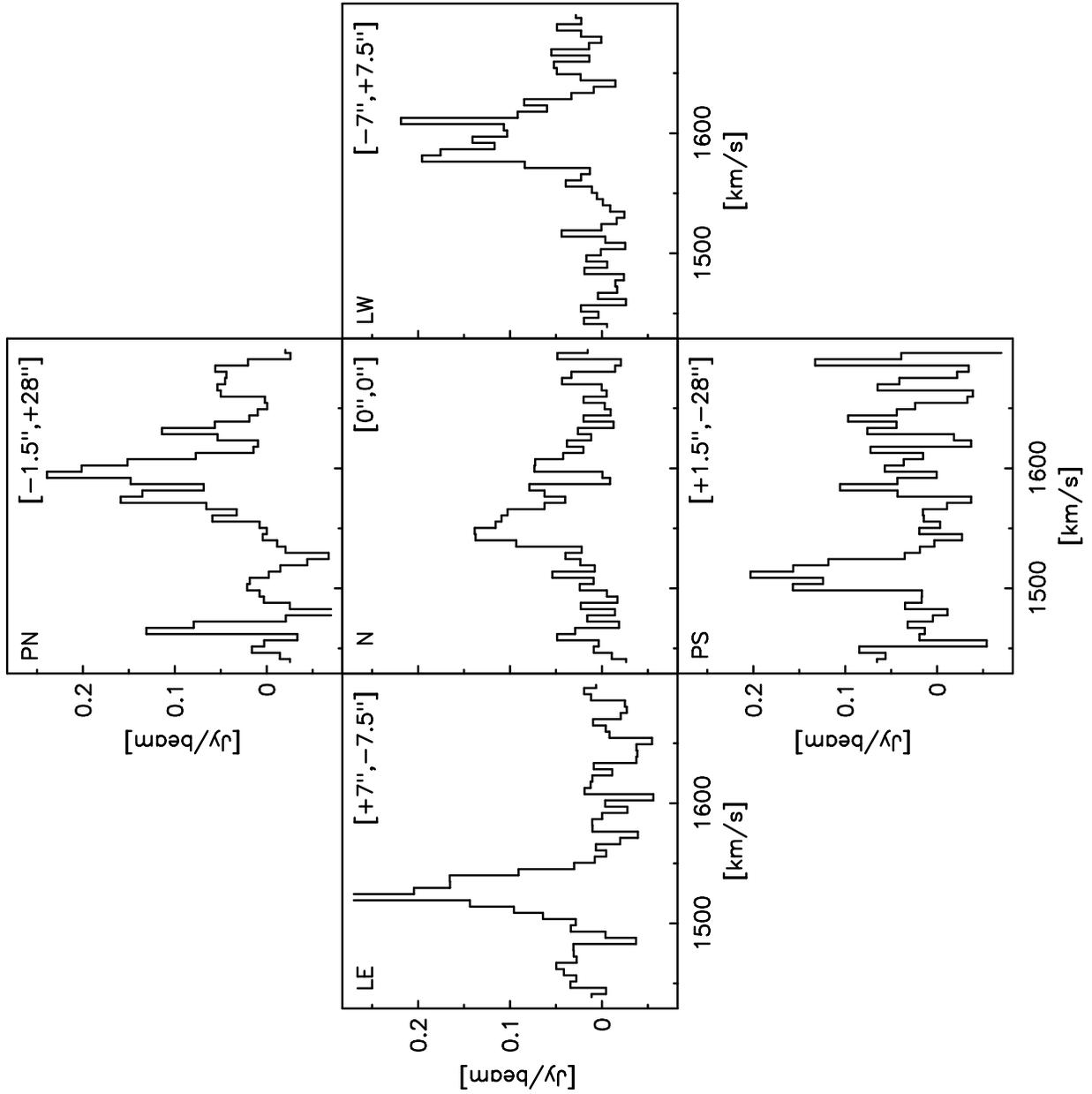}
\caption{
\COe\ line spectra extracted (from the primary beam corrected data) 
at the position of the nucleus, the emission peaks in the gas lanes,
and the emission peaks near the bar ends. The line center is clearly
shifting between the various positions. The positions of the spectra
are indicated in Fig. \ref{fig:COfet}.  }\label{fig:CObar}
\end{figure*}

\begin{figure*}[ht]
\epsscale{}
\plottwo{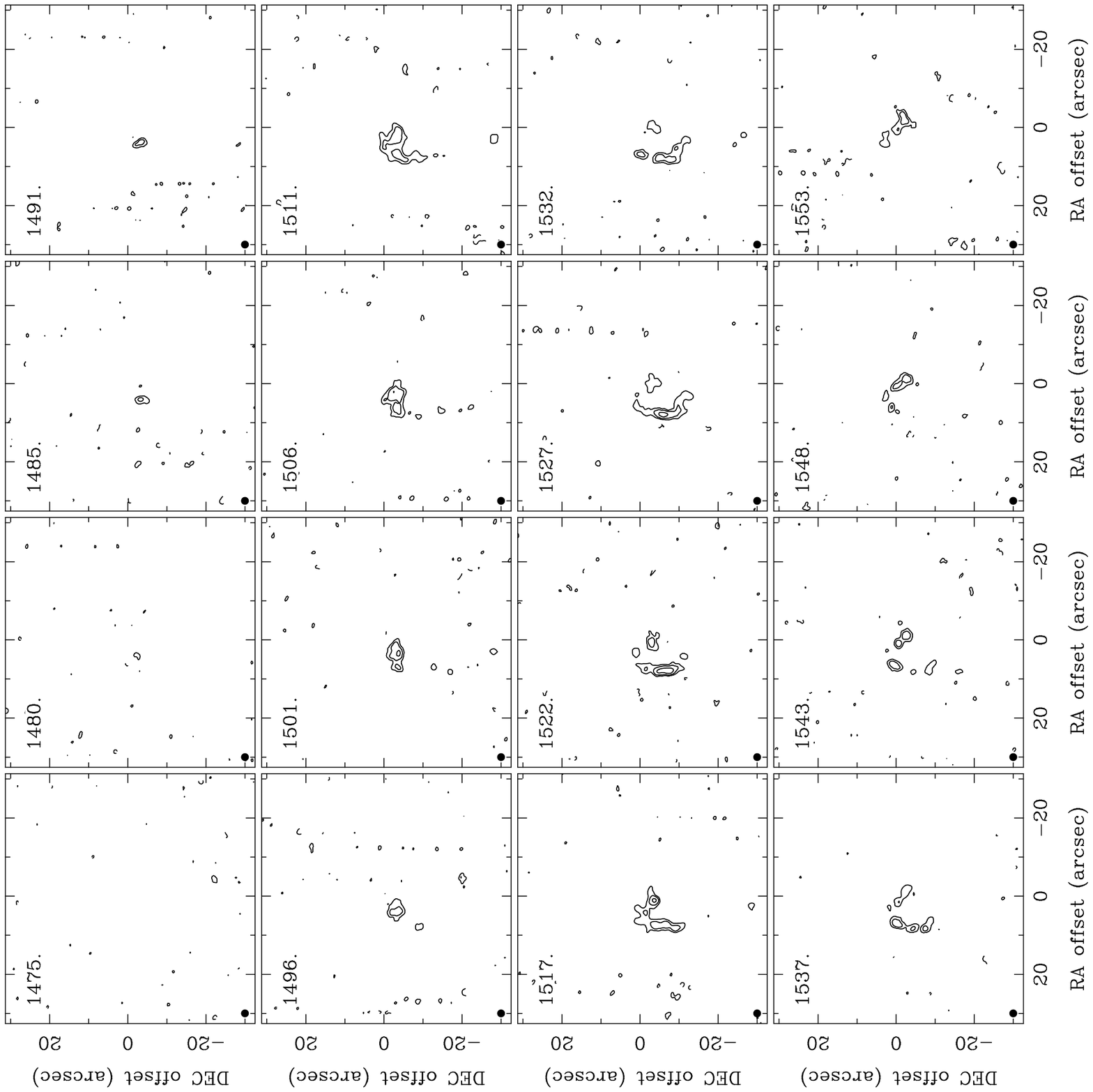}{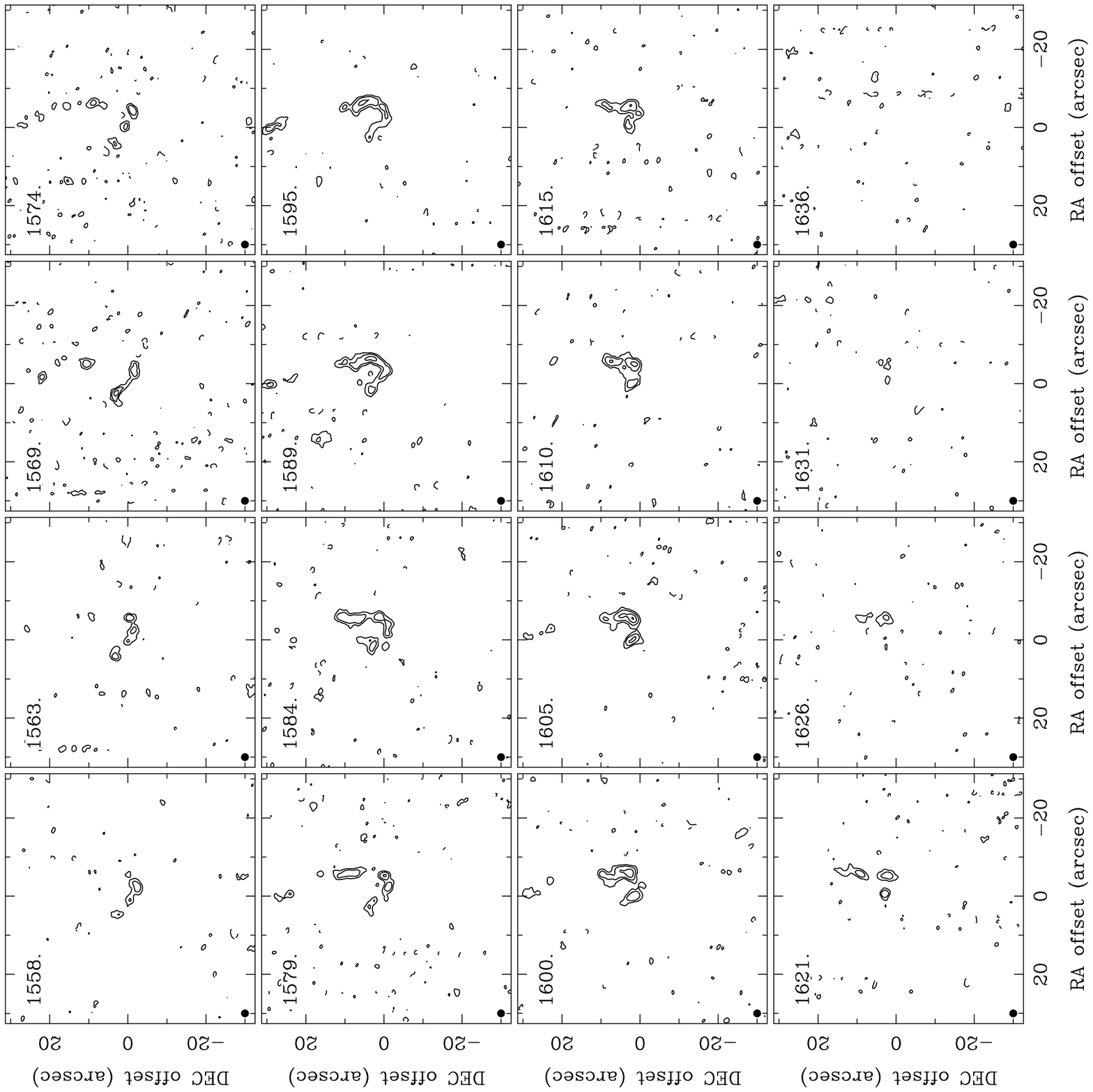}
\caption{
OVRO channel maps of the \COe\ line emission at a spatial resolution of
$2.0''$. The contours are at -3,3,5,8,11, and 14$\sigma$ with 1$\sigma
= 25$\,mJy\,beam$^{-1}$. The (LSR) velocity is given in each channel
map.  }\label{fig:COchan}
\end{figure*}

The \COe\ velocity field in the inner 12$''$ appears regular in most
locations, consistent with the ``spider'' diagram expected for a
differently rotating disk (Fig.~\ref{fig:COall} and
Fig.~\ref{fig:COchan}). The line of nodes changes at a radius of $\sim
7''$ between the inner region and the gas lanes indicating the
presence of non-circular motions. This can also be seen in the channel
maps (Fig.~\ref{fig:COchan}) where the molecular gas in the lanes
moves almost in north-south direction whereas in the inner $8''$ the
gas emission moves from south-east to the north-west. The distribution
of the velocity dispersion in the 2nd moment map is relatively smooth
with an average value of $\sim$ 10\,\kms throughout the observed
region (Fig.~\ref{fig:COall}).

\begin{figure*}[ht]
\epsscale{}
\plotone{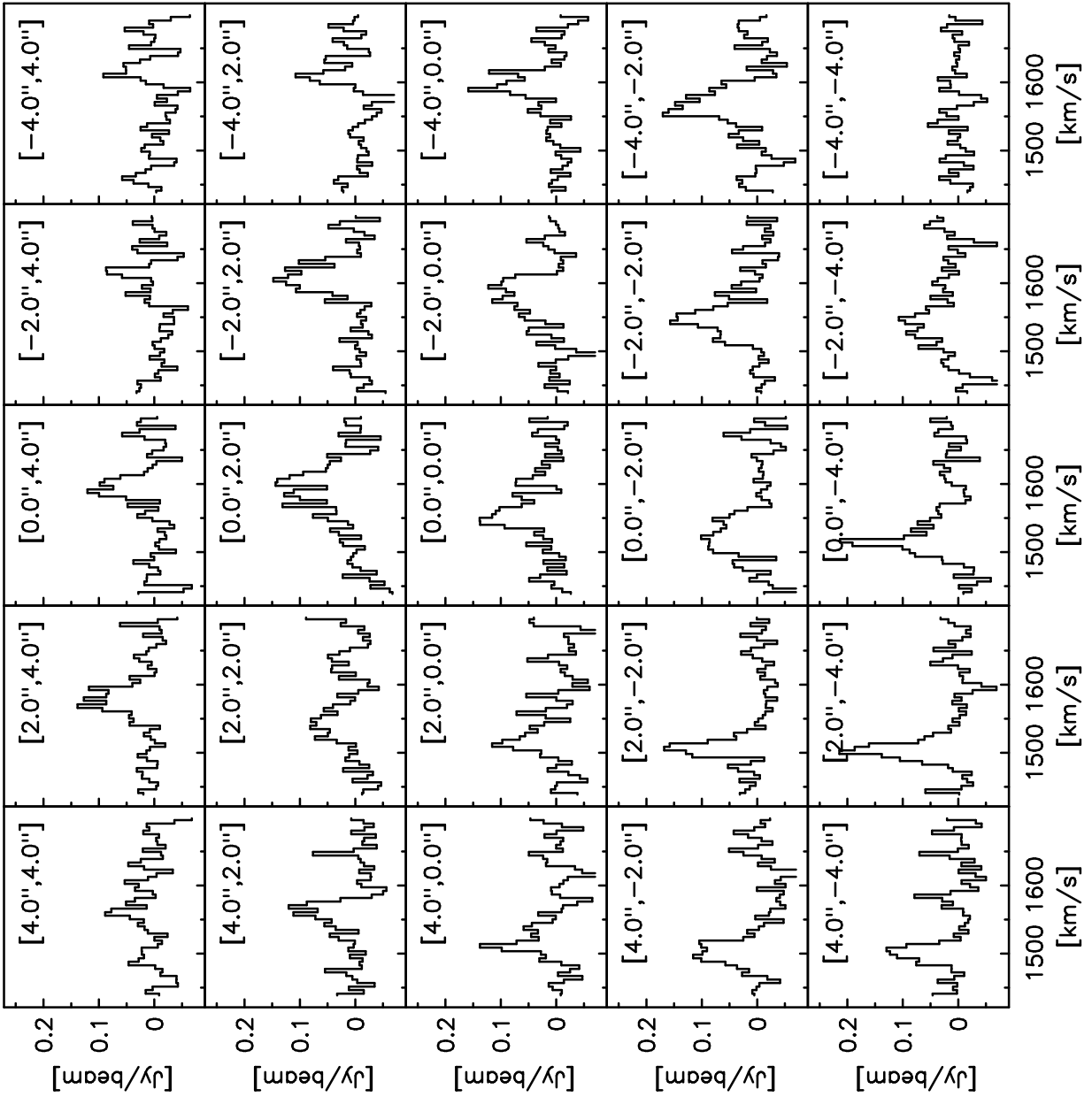}
\caption{
\COe\ line spectra extracted in the inner $8''$ in steps of $2''$. 
The spectra south-west of the nucleus show wider lines indicating the
presence of a second component.  }\label{fig:COnuc}
\end{figure*}

\begin{figure*}[ht]
\epsscale{.55}
\plotone{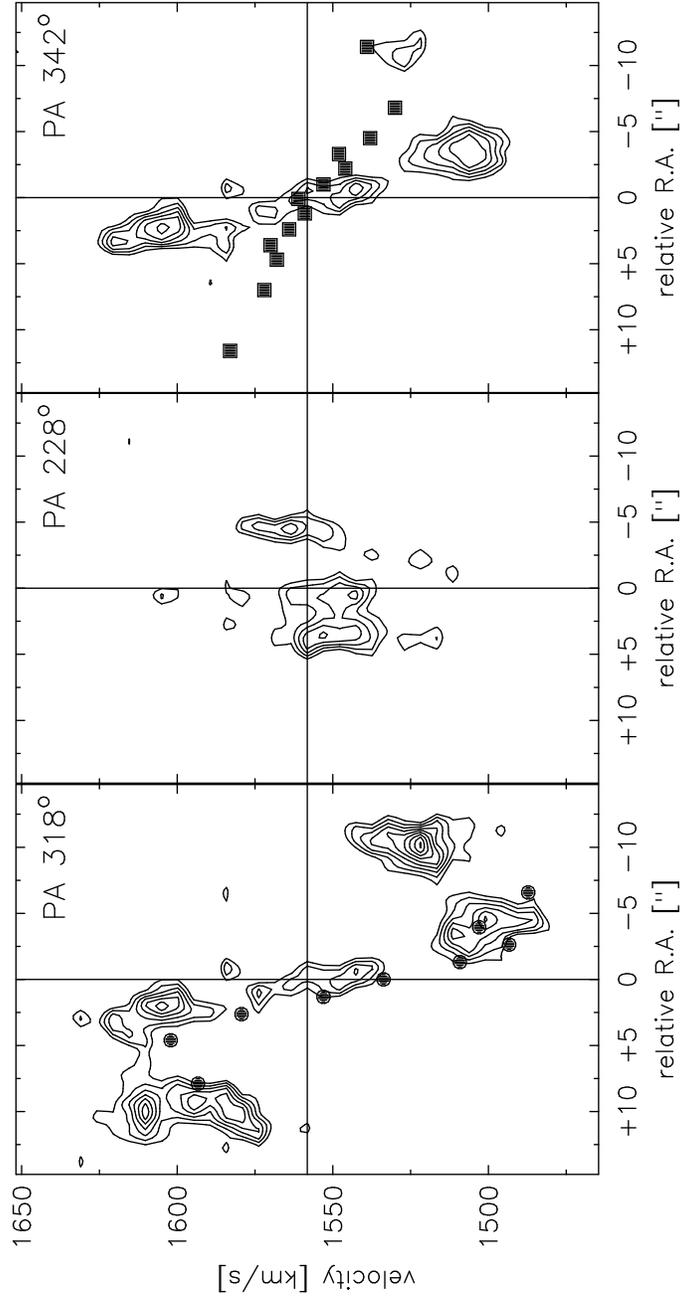}
\caption{ 
$pv$ diagrams along the major (PA 318$^{\circ}$; {\it left}) and minor
(PA 228$^{\circ}$, {\it middle}) kinematic axis of the \COe line
emission at a spatial resolution of $2.0''$ and a spectral resolution
of 5.2\,\kms. The filled points in the $pv$ diagram at PA
318$^{\circ}$ are the H$\alpha$ measurements of Rubin et
al. (1999). The $pv$ diagram at PA 342$^{\circ}$ ({\it right}) shows
in addition the stellar measurements of Heraudeau et al. (1998) in
filled squares. The contours start at 3$\sigma$ and are in steps of
1$\sigma$=25\,mJy\,beam$^{-1}$.  }\label{fig:pv}
\end{figure*}

The position-velocity ($pv$) diagram along the kinematic major axis
shows a steep rise in velocity to a distance of $r \sim 4''$. There is
good agreement with the H$\alpha$ $pv$ diagram for the same position
angle (Rubin et al. 1999; also Fig.~\ref{fig:pv}). The deviations are
probably due to the slightly different angular resolution and the
higher spectral resolution of the \CO\ data. Comparison with stellar
absorption line data along a position angle of PA\,$ = 342^{\circ}$
(Heraudeau et al. 1998) shows that the velocity of the stars is much
lower. It is unlikely that this is due to non-circular gas motions in
the inner $\sim 10''$, as the \CO\ velocity field
(Fig.~\ref{fig:COall}) as well as the H$\beta$ velocity field (Colina
\& Arribas 1999) are similar to a smoothly rotating disk with radial
motions of less than 10\,\kms\ (Colina \& Arribas 1999).  The lower
velocity of the stars in the central region very likely reflects the
fact that the stars have a different 3-dimensional distribution, and
therefore different kinematics. This is consistent with large
velocity dispersions of the stellar absorption lines ($\sim$
80\,\kms\ ; Heraudeau et al. 1998) compared to those of \CO\ gas
emission lines ($20-30$\,\kms; Fig. \ref{fig:COnuc}) in the nuclear
region. A similar behavior is seen in NGC~1068 where the derived
stellar rotation curve has lower velocities than the rotation curve
derived from the \CO\ data and the H$\alpha$ ~line emission
(Schinnerer et al. 2000a).

\subsection{Dynamical reference parameters}\label{subsec:dyn}

Before a rotation curve can be fit to the \CO\ data, the major axis
position angle and inclination, the dynamical center, and the systemic
velocity need to be determined. We used an iterative approach to
obtain these parameters by fitting tilted annuli to the observed \CO\
velocity field. For the fit we weighted points closer to the major
kinematic axis higher and neglected regions that lie within
15$^{\circ}$ of the minor axis. We used the routine `ROTCUR' in GIPSY,
in which only the rotation velocity and one parameter were fit at a
time, keeping all other parameters fixed.

\underline{\bf Position angle and inclination:} We estimate a
position angle of (317$\pm$5)$^{\circ}$ for the inner $10''$, where the
\CO\ velocity field should not be disturbed by the bar potential. This
agrees with Guhathakurta et al. (1988), who obtained a position angle
of (318$\pm$5)$^{\circ}$ from a fit to the large scale HI velocity
field, and with Colina \& Arribas (1999), who find a position angle of
$\sim(310 \pm 4)^{\circ}$ from modeling the H$\beta$ velocity field in
the inner $8''$. A robust determination of the inclination is
difficult because of the almost face-on aspect of NGC~4303 and the
optical disk disturbed due to tidal interaction with two nearby
companions (NGC~4292 and NGC~4303A; Binggeli et al. 1985). From the
optical images of Frei et al. (1996) using the lowest undisturbed
contours (excluding the regions of the spiral arms), we derive an
inclination of $\sim$ 25$^{\circ}$ by excluding radii that are already
affected by the bar which has a position angle of $\sim$
10$^{\circ}$. A fit to the \CO\ velocity field in the inner $20''$
gives an inclination of $i=(21\pm10)^{\circ}$. Cayatte et al. (1993)
find a similar value, and Ma, Peng \& Gu (1998) obtain 22$^{\circ}$
from a fit to the spiral structure. Note that Colina \& Arribas (1999)
used an inclination of $\sim 45^{\circ}$ assuming that the ionized gas
structure of the inner $8''$ is circular. However, as it seems
unlikely that the inner $8''$ disk has a different inclination, we
adopt the value for the outer disk of 25$^{\circ}$, which is
consistent with the inclination derived from the \CO\ velocity field,
and we use a position angle of 318$^{\circ}$ to fit the rotation
curve.

\underline{\bf The dynamical center:} Direct measurements of the
dynamical center from the \CO\ line data is complicated by the fact
that the line emission is not smoothly distributed in the inner
$8''$. Nevertheless, since the gas distribution appears
point-symmetric, the bar potential producing this pattern is plausibly
expected to be point-symmetric as well with respect to the dynamical
center. If we assume that the two gas lanes are symmetrically
distributed about the nucleus, we find that the dynamical center of
the galaxy is located within $1''$ of the central peak in the NICMOS
F160W image. (We verified this by aligning the dust lanes seen in the
HST F606W image and the gas lanes.) Thus the dynamical center is
identical with the peak of the old stellar population as seen in the
F160W image. We therefore adopt 12:21:54.99 +04:28:25.6 (J2000.0) as
the dynamical center for fitting of the rotation curve.

\underline{\bf Systemic velocity:} We find a systemic velocity of
$v_{\rm LSR} = (1559\pm3)$\,\kms\ using tilted ring fitting. This
translates into $v_{helio}=v_{\rm LSR}+9.5$\,\kms$=1569$\,\kms\ for
NGC~4303. Previous measurements in the HI line gave $v_{\rm helio} =
(1568\pm6)$\,\kms\ with the WSRT (Westerbork Synthesis Radio
Telescope) at an angular resolution of $10''$ (Warmels 1988), and
(1566$\pm$5)\,\kms\ with the VLA (Very Large Array) at an angular
resolution of $45''$ (Guhathakurta et al. 1988). Our value is in very
good agreement with the HI systemic velocity. The lower value of
$v_{\rm helio}=1552$\,\kms\ from the modeling of the H$\beta$ velocity
field (Colina \& Arribas 1999) might be due to the lower spectral
resolution of the H$\beta$ data. We used a value of $v_{\rm
LSR}=1560$\,\kms\ to fit the rotation curve.

\subsection{Rotation curve and dynamical mass}\label{subsec:rotcurve}

Since the molecular line emission in NGC~4303 is distributed very unevenly
throughout the disk, it is very difficult to obtain an optimally
averaged rotation curve. In addition, from the gas morphology we
expect strong non-circular motions in the straight gas
lanes. Therefore the rotation curve that we derive here is strongly
biased towards regions with higher emission, and is more
representative of the streaming motion outside the inner $10''$. On
the other hand, the coverage of the inner $10''$ is complete enough to
give a good estimate of the averaged rotation curve representing the
underlying gravitational potential.

\begin{figure*}[ht]
\epsscale{.70}
\plotone{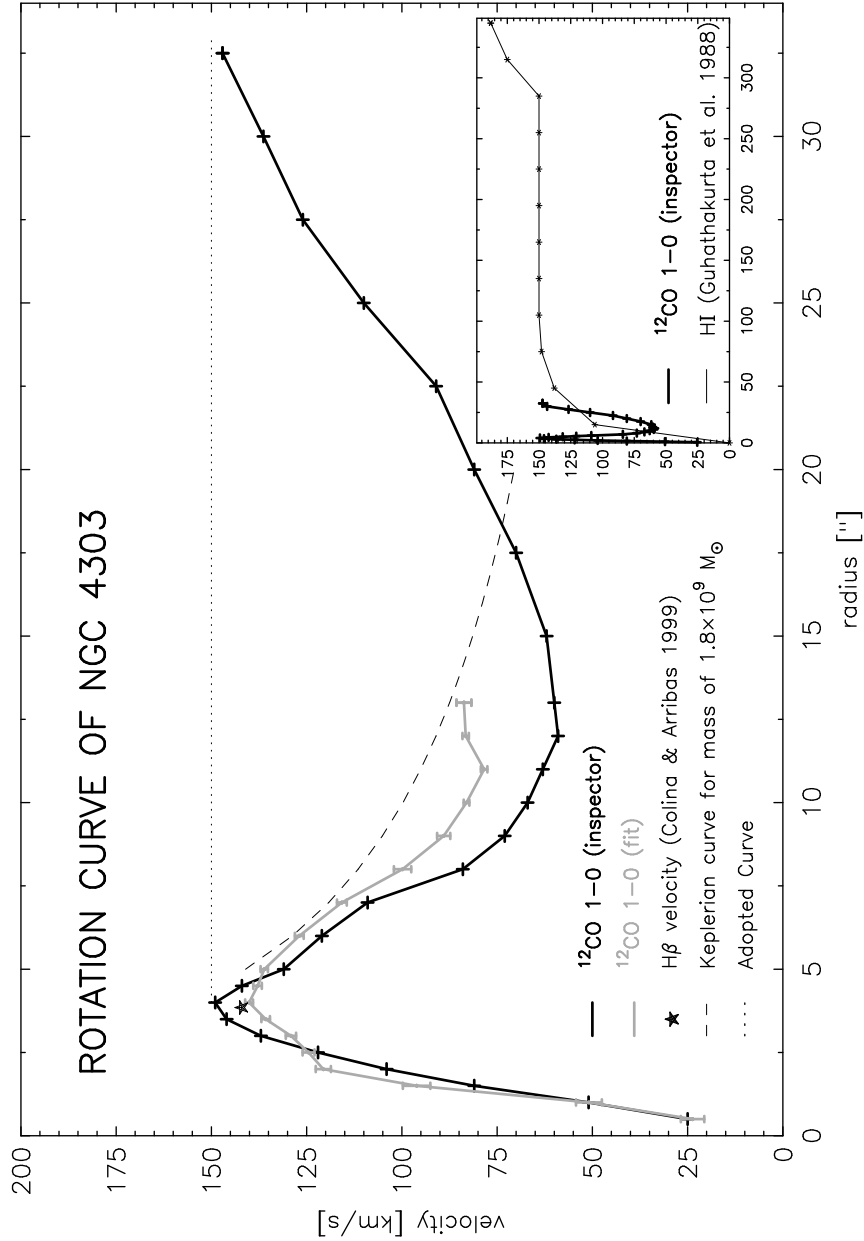}
\caption{
Derived rotation curve for NGC~4303. The results of the two different
fitting routines for the \COe\ data are shown in fat lines (grey:
ROTCUR; black: INSPECTOR). The derived rotation velocity of Colina \&
Arribas (1999) is given as well (star) after correcting for the
difference in inclination. The broken line represents the Keplerian
rotation velocity for a point mass of $1.8 \times 10^9$ \solm. The
molecular gas rotation curve is clearly sub-Keplerian for radii
between $r \sim 4'' - 18''$, strongly suggesting that non-circular
motions dominate the molecular gas motion here. Therefore, we adopted
a constant rotation velocity of 150\,\kms\ for these radii (dotted
line). The HI rotation curve of Guhathakurta et al. (1988) is shown
for comparison in the inset. See text for further discussion.
}\label{fig:rot}
\end{figure*}

Two methods were used to obtain the rotation curve
(Fig.~\ref{fig:rot}): A least square fit to the velocity field using
inclined rings (routine `ROTCUR') and a direct fit to $pv$ diagrams
along eight different position angles separated by steps of
22.5$^{\circ}$ for a good coverage of the galaxy disk (routine
`INSPECTOR'). The latter method allowed us to examine the effect of
beam smearing. In this procedure, a rotation curve is obtained by
fitting a tilted ring model to the whole data cube by comparing the
model with the $pv$ diagrams. This method is more robust against
multiple line components or additional non-circular components. It
also allows us to fit the rotation curve to the two outer \CO\
emission peaks at a radius of $r \sim 28''$.  As discussed above, the
dynamical center, the systemic velocity ($v_{\rm sys} = 1560$\,\kms),
the position angle (PA~$= 318^{\circ}$) and the inclination ($i =
25^{\circ}$) were kept fixed.

In the case of NGC~4303 the two rotation curves agree very well with
each other for the inner $r \leq 15''$ (Fig.~\ref{fig:rot}). The only
discrepancy is the peak velocity at a radius of $r \sim 4''$ which is
about 20 \kms\ higher for the rotation curve derived by `INSPECTOR.'
This is expected, since this method is less affected by beam
smearing. The derived velocity value at $r \sim 4''$ from the H$\beta$
velocity field (Colina \& Arribas 1999) is in very good agreement with
the \CO\ rotation curve when corrected for the different assumed
inclinations of the disk (see Section~\ref{subsec:dyn}). The drop of
the rotation velocity from $v(r=4'') \sim 150$\,\kms\ to $v(r=18'')
\sim 75$\,\kms\ is sub-Keplerian (Fig.~\ref{fig:rot}). Thus, the
derived rotation velocity can only be regarded as a lower limit. This
behavior confirms the presence of strong non-circular motion for radii
between $r \sim 4''-30''$. Therefore, we assume a constant rotation
velocity of 150\,\kms\ between $r=4''-30''$. For larger radii the
\CO, H$\alpha$ (Rubin et al. 1999) and HI (Cayatte et al. 1990)
rotation velocities are consistent.

\begin{figure*}[ht]
\epsscale{.55}
\plotone{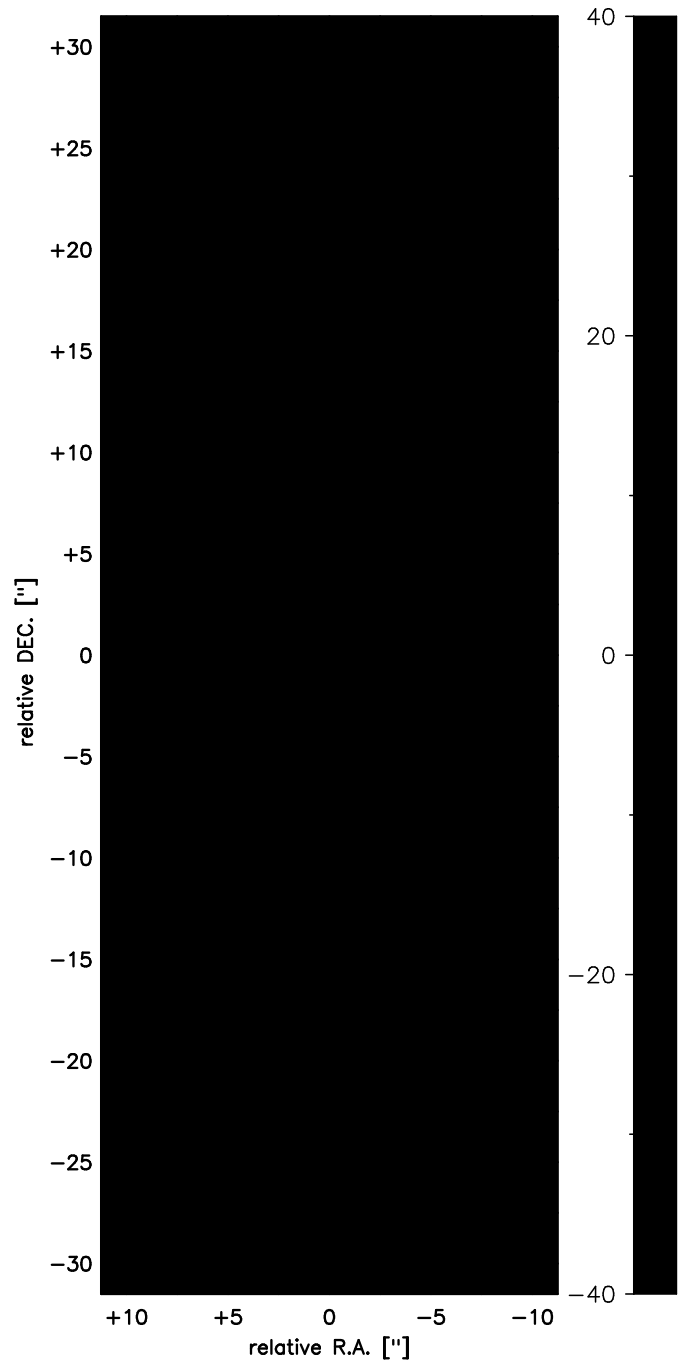}
\caption{
Residual velocity field. A model velocity field obtained from the
rotation curve was subtracted from the observed velocity field. Large
residuals are associated with the gas lanes indicating streaming
motion. Only small residuals are seen in the inner $8''$ demonstrating
that the assumption of a rotating disk is a good approximation.
}\label{fig:resvel}
\end{figure*}

We constructed a model velocity field from the derived rotation
curve. The residual velocity field of the observed minus the model
field (Fig.~\ref{fig:resvel}) shows differences of $\sim 7$\,\kms\
at the outer \CO\ peaks, and $\sim 5-10$\,\kms\ in the inner $12''$
where the \CO\ emission starts to form a spiral-like
structure. Therefore, the \CO\ line emission is well approximated by a
rotating disk with the derived rotation curve in the inner $12''$.
Outside that region, along the gas lanes, a maximum value for the
residual velocities of about $25-30$\,\kms\ is found at a radius of
$\sim 9.5''$ ($\approx 740$~pc) which is about the position of the
inner Lindblad resonance (ILR; see Section~\ref{sec:geomdyn}). 
After correcting for inclination, these values translate into $\sim$
60\,\kms\ which is the difference between the \CO\ rotation velocity
and the adopted rotation curve. If the spiral arms are trailing,
NGC~4303 is rotating clockwise and the north-eastern side is the near
side of the galaxy disk. As negative residuals are found in the
north-western gas lane and positive residuals in the south-eastern gas
lane this is consistent with strong inflowing motions, as predicted by
numerical simulations of gas flow in a barred potential (e.g. A92;
Piner et al. 1995; for a detailed explanation see also Regan, Sheth \&
Vogel 1999). This situation is similar to the findings of Regan et
al. (1999) for NGC~1530.

The dynamical mass $M(r)$ can be derived for the inner $8''$ where the
measured rotation curve agrees with solid body rotation using $v^2(r)
= G \frac{M(r)}{r}$ where $v(r)$ is the measured circular velocity
($G$ is the Gravitational constant). For $v(r=4'') \approx 150$\,\kms\ 
we get a dynamical mass $M_{dyn}(r=4'') \approx 1.6 \times 10^9$\solm.

\subsection{Molecular Gas Masses}

The \COe\ line fluxes $S_{CO}$ for several regions (nucleus, inner
disk, inner spiral arms, inner and outer gas lanes, and the outer
peaks as indicated in Fig.~\ref{fig:COfet}) are summarized in Table
\ref{tab:COfluxes}. The total line flux of $\sim 210$\,Jy\,\kms\ 
detected by the interferometer can be compared to the single dish flux
of $\int T_A^* d v = (11.9\pm2.1)$\,K\,\kms\ detected at the center
with the FCRAO 14m (Kenney \& Young 1988). Without any assumptions and
corrections for the size of the molecular gas disk, this translates
into $\sim 520$\,Jy\,\kms. Therefore the interferometer detects only
about 40\% of the single dish flux. Kenney \& Young (1988) give a
total \COe\ line flux of $(2280\pm470)$\,Jy\,\kms\ for the whole galaxy
using several single dish pointings, so that the molecular gas
emission mapped by the interferometer comprises only about 10\% of the
total molecular gas present. A recent BIMA SONG (Survey of Nearby
Galaxies) map of NGC~4303 with several pointings shows a large
fraction of the molecular gas along the outer spiral arms (Sheth
2001). This indicates that a large-scale disk of molecular gas is
present, and that the molecular gas in the spiral disk of NGC~4303 is
only partially redistributed towards the nucleus.

We used the \nhico\ conversion factor of $2 \times 10^{20}$
cm$^{-2}$\,(K\,\kms)$^{-1}$ from Strong et al. (1987) to derive
molecular gas masses. Note, that the value of the \nhico\ conversion
factor has an uncertainty of a factor of 2 - 3 (see e.g. Meier, Turner
\& Hurt 2000; Wei\ss ~et al. 2001). The molecular gas in the inner 
$8''$ ($\sim$ 630~pc) has a mass of about $M_{H_2} \sim 7\times10^7$
\solm ~(excluding the uncertainty of the conversion factor), which is
about $\sim$ 4\% of the dynamical mass of $\sim 1.6 \times 10^9$
\solm. Close to the nucleus about $10^7$ \solm\ of molecular 
gas are present. The amount of molecular gas mass in the two prominent
parts of the straight gas lanes is about twice the amount in the inner
630~pc.

\subsection{Molecular gas disk scale height} 

There are various ways to estimate the velocity dispersion in the
inner $8''$ of the molecular gas disk. The velocity dispersion is about
$10-15$ \kms\ in the 2nd moment map which can be regarded a lower limit
due to the 3$\sigma$ clipping which was used to derive the map. The
spectra in the nuclear region (Fig.~\ref{fig:COnuc}) show average line
widths of $15-30$~\kms\ which are a better measure of the observed
velocity dispersion. If the inner molecular gas `disk' is in
hydrostatic equilibrium, the velocity dispersion of the molecular gas,
after correcting for the rotational contribution due to the finite
beam size, can be used to derive a disk scale height. The observable
velocity gradient is $\sim 15$\,\kms~arcsec$^{-1}$ in the inner $8''$
(using the observed velocity of $v(r=4'') \sim 63$\,\kms). The true
velocity dispersion is then $15-20$\,\kms\ using quadratic
deconvolution to correct for the contribution from the rotation
velocity. Several theoretical relations between the velocity
dispersion and the disk height can be found in the literature. Quillen
et al. (1992; their equation 3.1) and Combes \& Becquaert (1997; third
equation in their introduction) have derived such relations for gas
disks in potentials similar to the ones of elliptical galaxies or
galaxy bulges. With the true velocity dispersion of $15-20$\,\kms, and
a rotational velocity $v(4.0'') \approx 150$\,\kms\ as well as the
associated molecular and dynamical masses, we find a disk thickness of
about $25-30$\,pc at a radius of $r \sim 4'' \approx 300$\,pc. This is
similar to the low molecular gas disk scale heights found in the central
few 10~pc in the two nearby Seyfert galaxies NGC~1068 and NGC~3227
(Schinnerer et al. 2000a,b).

\section{THE OBSERVED STELLAR AND DYNAMICAL PROPERTIES OF NGC~4303}\label{sec:mmstar}

\subsection{Dust extinction and star formation in the inner kiloparsec}
\label{sec:innerkpc}

\begin{figure*}[ht]
\epsscale{1.0}
\plotone{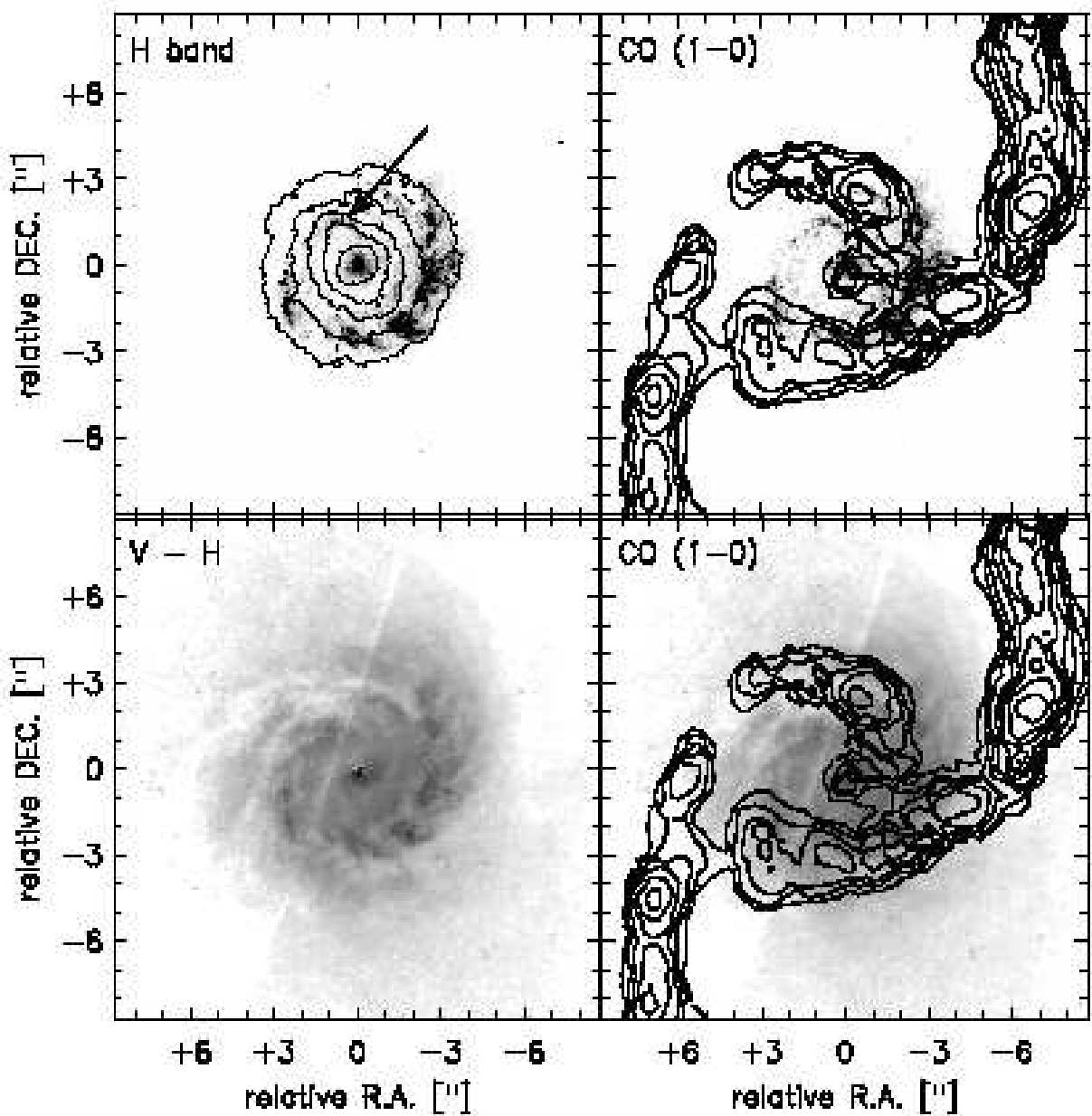}
\caption{
Comparison of the stellar clusters seen in the UV continuum
(gray-scale, {\it top}) to the inner bar seen in the HST $H$ band
(contours at 5\%, 10\%, 15\%, 20\% and 30\% of peak value, {\it top
left}), and the \CO\ line emission (contours at 20\%, 30\%, ... , 90\%
of peak value, {\it top right}). The extinction can be seen in the HST
$V-H$ map (blue colors are darker, {\it bottom left}) in comparison to
the \CO\ line emission (contours at 20\%, 30\%, ... , 90\% of peak
value, {\it bottom right}). The arrow in the {\it top left} panel
indicated the string of UV continuum which lies along the NIR bar and
is discussed in Section~\ref{subsec:gashst}.  }\label{fig:COhst}
\end{figure*}

In the inner kiloparsec, two or three dust lanes running from east to
north are present in the $V-H$ color map (Fig. \ref{fig:COhst}). The
eastern end of the outermost dust lane coincides with the northern gas
spiral arm seen in \CO\ at our resolution of 2''. Directly north of
the nucleus, the dust lane splits into two: a fainter part follows the
\CO\ pattern, but its much clearer extension continues into the
ring-like blue feature. This feature appears also in the UV continuum
as a string of young stellar clusters forming the starburst ring or
spiral at $r \sim 2.5''$ ($200$\,pc) (Fig.~\ref{fig:COhst}; see also
Fig. 1 of Colina et al. 1997). This continuation suggests that the
ring might be more extincted towards the northern part, or that the
star formation is weaker in the north of the ring, and ceases at the
dust lane (see also section~\ref{subsec:gashst}). In a second
scenario, the northern molecular spiral continues inwards, and is seen
as a dust lane, with star formation starting at still smaller
radii. It is interesting to note that if the UV morphology is
interpreted as a spiral, it is a {\it single-arm} spiral originating
from the northern molecular spiral arm. The young stellar clusters are
offset from the molecular gas distribution next to the disturbed,
southern molecular spiral arm.  This arm discontinues in \CO\ at
PA$\sim135^o$, and finds no continuation in other features, aside
possibly for the second eastern dust lane, which appears almost
straight in the V-H color map. It appears that Apparently, the massive
star formation next to the southern molecular arm has affected the gas
distribution in that region (see Section~\ref{subsec:sfmods}).

Colina \& Wada (2000) derive an excess extinction of about 1.7$^{mag}$
for the northern dust lane relative to values for face-on spiral
galaxies, similar to values observed in the circumnuclear regions of
other spiral galaxies (e.g. Regan \& Mulchaey 1999; Martini \& Pogge
1999). This translates into an extinction of about 20$^{mag}$ in the
UV band using the effective extinction law derived for the stellar
continuum in starburst galaxies (Calzetti, Kinney \& Storchi-Bergmann
1994). This high extinction in the northern dust lane would be enough
to obscure any UV cluster in this region, thus it seems possible that
the star formation might actually occur in the ring traced by the
UV clusters and the prominent (northern) dust lane.


Comparison with the optical line, and the optical and NIR continuum emission
suggests that the stellar clusters are between $5-25$ Myr old and have
stellar masses of about $(0.5-1.0) \times 10^5$\,\solm\ (Colina \&
Wada 2000). Colina \& Wada (2000) noted that the UV knots east of the
inner bar are older ($\sim 10-25$\,Myr) and more massive (up to $2.3
\times 10^5$ \solm) than the ones west of the NIR bar ($\sim
2.5-7.5$\,Myr), suggesting an age trend within the star forming ring.
The molecular spiral arms can be divided into about six individual
clumps with molecular gas masses of $M_{H_2} \sim 7 \times 10^6$
\solm\ and sizes between $\leq 2''$ and $3''$ ($\leq 150-230$\,pc), 
typical for GMC complexes (the 0th moment map is sensitive to cloud
masses above 10$^6$ \solm).  The star formation efficiency (SFE) for a
molecular cloud lies typically between 3\% (for the total cloud
including low-mass star formation) and 24\% (for the hot cores; see
review by Elmegreen et al. 1999). If one allows for an elevated SFE of
about 10\% in the circumnuclear region, which seems plausible given
the different physical properties compared to the disk of galaxies,
then a single massive GMC complex similar to those observed in
NGC~4303 could be a sufficient to form a dozen stellar clusters with
an average stellar mass of $0.7 \times 10^5$ \solm\ (Colina \& Wada
2000) as inferred from the UV continuum (see section
\ref{subsec:sfmods}).

Colina \& Arribas (1999) isolated a second velocity component in the
[O~III] line which has a velocity field almost perpendicular to that
of the molecular gas. They interpret this as the signature of an
ionization cone which is observed in a number of AGNs (e.g. Schmitt
\& Kinney 1996). Since the north-east side is the near side of the
galaxy, the ionization cone is in front of the south-western part of
the galaxy disk. The obvious asymmetry in the extinction in the inner
$8''$ could be then partly due to the presence of an ionization cone
most affecting the region south-west of the nucleus. This suggests
that the circumnuclear colors in NGC~4303 are (a) strongly modified by
the varying colors of the different stellar populations (the clusters
are obvious in the $V-H$ map), and that (b) an ionization cone might
affect the overall distribution of the extincting material. This
cautions the interpretation of dust features seen in HST color maps as
being only due to molecular gas condensations.

\subsection{The bar properties}\label{subsec:barprop}

The large-scale bar in NGC~4303 runs almost north-south (PA $\sim
10^o$) inside the spiral arms. Martin (1995) and Chapelon et
al. (1999) find its semi-major axis length to be about $a \approx
20''$. Laine et al. (2002) revised this result and obtained a
deprojected semi-major axis length of $a \approx 47''$ ($\approx
3.5$\,kpc) and a deprojected axial ratio of $b/a \sim 0.34$. Because
of the large discrepancy between these results, we fit ellipses to the
$gri$ images of Frei et al. (1996) and find a semi-major axis length
between $30''$ and $40''$. A more exact determination of the scale
length is hampered by the resolution of the Frei et al. data and the
fact that the spiral arms start at a radius of $\sim 40''$, which must
be close to the end of the bar. Since our value for the semi-major
axis appears to be more consistent with the dynamical models where the
peak gas density is close to the end of the bar (see
Section~\ref{sec:geomdyn}), we assume a deprojected semi-major axis
length of $30'' \leq a \leq 40''$. With its axial ratio the bar is
considered a strong bar, as defined by Friedli \& Martinet (1993). The
large-scale spiral arms are sites of massive star formation as can be
seen in the H$\alpha$ line emission (Banfi et al. 1993, Koopmann,
Kenney \& Young 2001). However, no prominent H$\alpha$ line emission
is found along the bar. This is in agreement with model predictions
that the shear along {\it strong} bars can be too high for stars to
form.

The secondary bar with a deprojected semi-major axis length of $\sim
2.2''$ ($170$\,pc) is visible in the HST $H$ band image inside the
stellar clusters (Colina \& Wada 2000; also Fig.~\ref{fig:COhst}).
The deprojected axial ratio of the inner bar of $b/a \sim 0.73$
(Perez-Ramirez et al. 2000) is at the high end of observed ratios for
secondary bars (Jungwiert, Combes \& Axon 1997).  If the observed
axial ratio is not diluted by the underlying bulge component it
indicates a weaker bar potential for the secondary bar relative to the
primary bar.  A string of faint star-forming clusters can be seen in
the UV image north of the nucleus (e.g. Fig.~\ref{fig:COhst}; Colina
\& Arribas 1999). It coincides with the leading side of the secondary
NIR bar assuming it has the same rotation direction as the
outer/primary bar.

The ratio between the semi-major axes of the two bars is about 15, 
which is at the upper end of the observed range (Jungwiert, Combes 
\& Axon 1997; Maciejewski \&
Sparke 2000; Laine et al. 2002). The deprojected position angle of the
secondary bar seen in the HST $H$ band image relative to the kinematic
major axis is PA\,$\approx 80^{\circ}$. Therefore the two stellar bars
are oriented at about $\approx 30^{\circ}$ with respect to each other
in the plane of the host galaxy.

\subsection{Position of dynamical resonances in NGC~4303}\label{sec:ilr}

Assuming the standard relation of $r_{CR} \approx 1.2 a$ (e.g. A92)
between the bar semi-major axis length $a \approx 30''-40''$ and
corotation radius $r_{CR}$ where the angular velocity $\Omega$ is
equal to the bar pattern speed $\Omega_P$, we find that the corotation
radius lies between $36''$ and $48''$. Thus the primary bar rotates
with a pattern speed of about $\Omega_P \sim 40-53$\,\kms\,kpc$^{-1}$.

For this pattern speed, and the rotation curve derived in
Section~\ref{subsec:rotcurve} (see Fig~\ref{fig:rot}), the large-scale
bar has two ILRs. The inner ILR (iILR), when localized directly from
the data, lies at $\sim 2.0''$ radius, and coincides with the
semi-major axis radius of the nuclear bar ($2.2''$; Perez-Ramirez et
al. 2000).  This puts the corotation resonance of the inner bar
between the two ILRs of the outer bar, roughly in agreement with
theoretical expectations (e.g. Friedli \& Martinet 1993, Maciejewski
\& Sparke 1997). However, the dynamically preferred scenario
(Maciejewski \& Sparke 2000) requires that the inner bar ends far
outside the iILR of the main bar, so there is a sufficient set of
orbits that can support it. We discuss this discrepancy with the
observed near-overlapping of one bar's end with the others iILR in
Section~\ref{subsec:gashst}.

The position of the outer ILR (oILR) can be determined using the
adopted flat rotation curve for this radial distance as an upper
measured limit (see Section~\ref{subsec:dyn}). We derive 
the oILR to be at $10''-14''$ radius.

\section{MOLECULAR DATA AND THE DYNAMICAL MODELS}\label{sec:geomdyn}

\subsection{Outer Regions: The Overall Geometry}

The gas response strongly depends on the gravitational potential, and 
especially on the properties of the stellar bar component (A92). A
large variety of possible gas morphologies is explored by A92.
The overall distribution of the \CO\ line emission in the inner
arcminute of NGC~4303 is very reminiscent of the standard model~001 by
A92. The molecular gas is concentrated towards the nucleus in two
narrow straight gas lanes, and in additional emission peaks close to each
end of the bar. Near the nucleus the straight lanes start to curl inward
resembling a structure similar to a `nuclear spiral' or `nuclear
ring'. As pointed out by A92 such behavior can only be observed if
an inner Lindblad resonance (ILR) is present. The observed structure
can be explained by the gas moving gradually from $x_1$ orbits (the
family of stable periodic orbits elongated along the bar major axis)
to $x_2$ orbits (the family of stable periodic orbits elongated along
the minor bar axis; in the notation of Contopoulos 1981), which only exist
in the presence of an ILR.

The amazing similarity between the A92 standard model and the \CO\
distribution in NGC~4303 suggests a bar strength consistent with the
model axial ratio of about 0.4. This roughly agrees with the
deprojected ratio of $b/a \sim 0.34$ of Laine et al. (2002). Note
though that the observed value is integrated over the total light that
includes the bulge, while the models give the axial ratio of the bar
only. Thus the observations suggest that the axial ratio of the bar
may be considerably lower than 0.4. Also, the observed angle between
the straight part of the gas lanes and the bar major axis is $\sim
10^{\circ}$ in NGC~4303 which is half the value found for model~001
(A92). This reinforces the supposition that the actual ratio of the
bar might be even lower than 0.4.

If the northern and southern emission peaks at a distance of $\sim \pm
28''$ are due to the dynamics imposed by the large-scale bar, they
should lie inside the bar potential, roughly at the position of the
4:1 resonance (Englmaier \& Gerhard 1997; see also Fig.~2(d) of
A92). For a flat rotation curve, the 4:1 resonance occurs at $0.65$ of
the corotation radius, which places the corotation at $43''$ in this
interpretation. This is consistent with the value that we assumed, and
after dividing by $1.2$ it gives the bar's semi-major axis of $36''$.

The qualitative comparison between the A92 standard model~001 and the
distribution of the \CO\ emission in NGC~4303 shows very good
agreement. The fact that some H$\alpha$ emission is associated with
the two outermost emission peaks (e.g Fig. 3.3 in Sheth 2001, Fig. 5
in Koopmann, Kenney \& Young 2001) is also in agreement with the model
that has a factor of $6-10$ lower shear in these regions.  This is
sufficiently low to allow star formation to occur (A92). The
higher-resolution models of Maciejewski et al. (2002) clearly show
that the gas density peaks correspond to an almost shear-free region
along the bar at about the position of the 4:1 resonance. The
resolution of the A92 models is, unfortunately, too coarse to allow
for a direct comparison of the dynamics in the (circum)nuclear region
itself.

\subsection{\CO\ in the inner Kiloparsec}\label{sec:dyncomp}  

To date, only a few high-resolution dynamical calculations have been
done to investigate the gas flow in the inner kiloparsec of barred
galaxies. Englmaier \& Shlosman (2000) studied gas response in nuclear
regions of a barred potential similar to that in the standard model of
A92. They found that the velocity dispersion in the gas (which is
assumed to be isothermal) together with the potential shape determines
whether the gas settles on a nuclear ring, disk or spiral.  In
particular, high-velocity-dispersion gas forms a nuclear grand design
spiral inside the ILR. This structure can be explained in terms of
gas-density-wave theory as a response of non-selfgravitating gas to
the torque of the bar. A gas density wave can exist inside the ILR, in
a region where the torque is not too strong (the $x_2$ orbits are
almost circular), and where no shocks are present. The pitch angle of
the nuclear spiral has a strong dependence on the gas velocity
dispersion, and is higher for higher sound speed, which is the measure
of velocity dispersion. Maciejewski et al. (2002) noticed that
although weak waves inside the ILR can be described by linear-wave
theory, sometimes the strong straight shock in the large-scale bar can
propagate all the way towards the center as a spiral shock, rendering
the linear wave approximation invalid. Strong streaming motions seen
along the gas spirals in NGC~5248 (Jogee et al. 2002) may be
indicative of such a spiral shock there.

Since the nuclear molecular gas emission in NGC~4303 shows a
spiral-like pattern in the inner $8''$, a detailed comparison between
the \CO\ data and the predictions of the dynamical models by
Englmaier \& Shlosman (2000) was performed in order to assess how far
inwards the outer bar dominates the gas dynamics. The agreement
between model predictions and the data is surprisingly good given the
fact that no tailored models with the exact galaxy parameters are
used.

\underline{\bf Rotation curve:} The azimuthal velocity of the molecular
gas decreases in the region where the shock is strongest, i.e. close
to the ILR, creating a net inflow. It is very reminiscent of Fig.~8 of
Englmaier \& Shlosman (2000). Although that figure displays velocities
at the maximal gas condensation, our derived \CO\ rotation curve of
NGC~4303 is also strongly biased towards bright emission, and thus can
be explained by inflow in a strong bar in the context of the
models. Comparing our curve to the model one, we find that inside a
radius of $r \sim 4''$ where the rotation curve is rising, the \CO\
rotation curve can be used as a tracer for the dynamical
mass. According to the model, the (outer) ILR coincides with the
largest difference between the rotation curve and the azimuthal
velocity. Since there are no other measurements of the
rotation velocity in NGC~4303 with sufficient angular resolution, the
oILR can be placed at a radius between $10''$ and $15''$ where our
measured \CO\ rotation velocity is lowest. This location agrees well
with the estimate in Section~\ref{subsec:barprop}, for which we
assumed a constant rotation speed.

\underline{\bf Density distribution:} The models predict that the
arm-interarm contrast in the nuclear spiral is much smaller than the
contrast between the straight shock and the surrounding disk (see
Fig.~4 of Englmaier \& Shlosman 2000). A similar behavior is observed
in NGC~4303 (Fig.~\ref{fig:damp}). The arm-interarm density contrast
{\em inside} the transition radius $R_t \approx 5''$ is about two in
NGC~4303, as is found for the model\footnote{Englmaier \& Shlosman
(2000) define the radius where the shock front crosses the bar major
axis for the first time as the transition radius $R_t$. Below this
radius the gas density wave is driving the gas structure and forms a
grand-design spiral structure.}. However, the contrast between the
straight outer shock and the outer disk in NGC~4303 is about three
times higher than in the model. As noted by Englmaier \& Shlosman
(2000) their plot shows only a lower limit as the outer shocks are
unresolved in their model. The gas lanes in NGC~4303 are unresolved in
our \CO\ data, implying that they must be significantly narrower than
$150$\,pc ($2''$). The dust feature in the HST $V$ band image suggests
a width of about $1.5''$ ($120$\,pc). This implies that the contrast
should be even higher. On the other hand, for deriving the
arm-interarm contrast we assumed that there is no low-level smooth
large-scale emission present between the gas lanes. Since we detect
only about 40\% of the single dish flux, this assumption is not
completely correct. For example, the effect of such a smooth component
at the 1$\sigma$ level would be to lower the contrast by about a
factor of three for larger radii. Therefore, the measured arm-interarm
contrast is very likely a good approximation. A factor of 1.5 higher
velocity dispersion is found for the density peaks within the gas
lanes compared to the average dispersion of $10$\,\kms\ in the rest of
the gas lanes (Fig.~\ref{fig:COall}), which might be due to the high
shear in these regions. This may imply that the velocity dispersion is
higher in the denser regions past the shock, and may put constraints
on the post-shock star formation, e.g. may limit cluster sizes.  In
contrast to other barred galaxies almost no H$\alpha$ emission is seen
along the bar (e.g. Sheth 2001), despite the high molecular gas
density.

\underline{\bf Pitch angle:} The behavior of the pitch angle in a
spiral arm can be a clear diagnostic for the presence of a gas density
wave (Englmaier \& Shlosman 2000). In our case, the gas pitch angle
drops for radii between the ILR and the transition radius $R_t$,
presumably due to the transition from $x_1$ to $x_2$ orbits
(Fig.~\ref{fig:pitch}). Inside the transition radius the pitch angle
rises again, indicating that another mechanism --- a gas density wave
--- is responsible for the spiral structure. Measuring the pitch angle
is complicated by the fact that the spiral structure in the \CO\ line
emission is not continuous. In addition the southern nuclear spiral
shows distortions which are probably related to the ongoing star
formation in the UV ring (see Section~\ref{sec:mmstar}). In order to
measure the pitch angle, the $0^{th}$ moment map was deconvolved using
the LUCY algorithm with 1000 iterations and deprojected
afterwards. The pitch angle was measured by eye fitting a tangent to
the spiral structure at several radii. To verify the result, the
spiral structure seen in the \CO\ emission is locally fitted by
various logarithmic spirals with a constant pitch angle ($40^{\circ},
20^{\circ}$ and $45^{\circ}$) in Fig.~\ref{fig:logsp}. The change of
the pitch angle close to the transition radius $R_t \approx 5''$ is
quite obvious. There is an indication that the pitch angle is rising
again after its drop at the transition radius, as expected from the
model of a nuclear gas density wave. Note that in Fig.~\ref{fig:pitch}
we show the pitch angle of the northern spiral only, and its
morphology strongly deviates from the dust and UV continuum morphology
inside of the transition radius (see Section \ref{sec:innerkpc}). It
bends almost directly towards the nucleus, while the dust and UV
continuum appear to form a ring.  We will return to this discrepancy
in Section \ref{subsec:gashst}.

Assuming that the indicated rise is real, the measured value of the
pitch angle is higher than in the model by Englmaier \& Shlosman
(2000). It may be due to the fact that the inner spiral is beyond the
linear regime explored in that model. The primary bar of NGC~4303 is 
stronger than the one in the model, and it may generate a stronger 
large-scale shock, which even in the central parts cannot
be approximated by the linear wave theory, as in the spiral-shock
model by Maciejewski et al. (2002).

\section{LINKING \CO\ TO HST DATA: DYNAMICS AND MORPHOLOGY 
IN THE INNERMOST 500 PC}\label{subsec:gashst}

As shown in the previous section, Englmaier \& Shlosman's (2000)
models of gas flow in a single bar are in a good agreement with the
observed molecular gas behavior in the inner kiloparsec of NGC~4303
down to the 2-arcsec beam size of our data. The dynamics in this
region can therefore be already well described by a model of a single
bar with an ILR. In particular, we do not detect any strong inflowing
motions throughout the whole inner bar, which would be characteristic
for a {\it gaseous} inner bar as postulated by Shlosman et al. (1989).
However, a secondary or inner, likely stellar, bar of $\sim 4''$ length is
clearly present in the HST $H$ band image (e.g Fig. \ref{fig:COhst}).
Due to its small size the use of our \CO\ data to search for inflow
inside the inner bar is limited. Nevertheless, we can use the
morphological information of the high resolution HST images ($\sim
0.2''$) to shed more light on the smaller-scale gas morphology in the
presence of the secondary bar.

In the inner 4'' the \CO\ morphology diverges from that seen in UV and
traced by dust lanes (see section \ref{sec:innerkpc}). To examine the
behavior of the three tracers (CO, UV, and dust), we plot the distance
from the nucleus as a function of the position angle in the inner
kiloparsec for all of them (Fig.~\ref{fig:pitch}). For PA = 330$^o$
and 345$^o$ the parameters of the two dust lanes (described in
Section~\ref{sec:innerkpc}), are given. The weaker dust lane follows
the \CO\ pattern, while the more prominent lane
joins the nuclear UV ring at PA = 0$^o$. The UV continuum
forms almost a ring, as it is evident from the constant radius. Its
geometry only deviates from a ring for position angles of $\geq
250^o$, where the string of UV-emitting regions seems to curl
towards the center. The derived UV pitch angle oscillates around
20$^o$, and does not show the rapid rise inwards as observed for the
pitch angle of the \CO\ spiral.

Two interpretations of the UV images and the $V-H$ color maps seem
plausible, and are discussed in detail below: (A) The UV continuum
forms a nuclear ring which is extincted north of the nucleus. This
ring could then be related to gas dynamics expected in a double
barred system. (B) The UV continuum forms a single-arm spiral. In
this scenario, the inner bar might not at all affect the gas dynamics.
In order to distinguish between the two scenarios, kinematic data of 
higher angular resolution are crucial.

\subsection{Scenario (A): A nuclear ring}

The outer UV continuum sources form a ring, onto which the \CO\ spiral
converges, as seen in the transition from \CO\ (PA $\geq 200^o$) through
dust ($250^o \leq$ PA $\leq 360^o$) to UV (PA $\geq 0^o$) in
Fig.~\ref{fig:pitch}. Here we show that the ring morphology is consistent 
with gas flow around a dynamically possible {\it stellar} secondary bar.

\underline{\bf Dynamical constraints on double barred galaxies:}
A fixed relative position of two stellar bars cannot be stable
for a long time, unless they remain parallel or orthogonal to each
other (see e.g. Friedli 1996). The deprojected relative angle between
the two bars in NGC~4303 is $\sim 30^{\circ}$, and thus the bars most
likely do rotate with respect to each other. In numerical simulations
of Friedli \& Martinet (1993) and Rautiainen \& Salo (1999) the two
bars decouple dynamically, and may survive for long times,
so that they can be treated as stable, long-lived systems. The
presence of a moderate ILR of the primary bar is essential for the
decoupling. Maciejewski \& Sparke (2000) performed a search for orbits
supporting bars within bars, and found that orbits supporting the
inner bar originate from the $x_2$ orbits of the outer bar, and
therefore the inner bar should form inside the ILR of the outer one.
Maciejewski \& Sparke (2000) were the first ones who postulated 
that straight shocks or dust lanes in the secondary bar are unlikely, 
because the orbital structure forces the secondary bar to end well 
inside its corotation. In such a bar, oval rings are expected instead 
of straight shocks. This finding was confirmed by hydrodynamic 
modeling by Maciejewski et al. (2002), which shows that the main
effect of the secondary bar is to widen the nuclear ring created by
the single bar. This ring evolves into an ellipse which rotates with
the inner bar.

\underline{\bf The morphology in the inner 500~pc:}
The distance between the straight gas lanes in the primary bar of
NGC~4303 is much larger than the size of the inner bar, which differs
from the modeled setup (Maciejewski et al. 2002). This allows the
molecular gas spiral generated by the large-scale shock in the main
bar to propagate considerably inwards before entering the region
dominated by the secondary bar. Contrary to the model, the straight
gas lanes in the main bar reach directly to the region dominated by
the secondary bar. This difference may explain why the molecular gas
in NGC~4303 forms a spiral outside the inner bar. The innermost
morphology, where the UV continuum forms a ring, is consistent with
the model that predicts no straight shocks along the inner
bar. However, one observed feature contradicts the model: a string of
UV regions, which lies almost north of the nucleus (indicated by the
arrow in Fig.~\ref{fig:COhst}; see also Fig. 2 of Colina \& Wada
2000), coincides with the leading edge of the northern half of the
secondary bar. These UV regions may indicate a shock along the inner
bar that is not prohibitive to star formation. Thus, this secondary
bar which is smaller than the one in the models of Maciejewski et
al. (2002) can not only form a ring around itself, but it may also
generate gas flows similar to those seen in the main bar, and,
therefore, it might enhance gas inflow to the galactic center along
the string of UV regions. If this is so, it is baffling to see no
counterpart of this string on the other side of the bar, especially
since the UV regions in the ring are more profound south of the
nucleus.

\underline{\bf Does NGC~4303 have an inner ILR?}
In Maciejewski \& Sparke's (2000) model, the inner bar cannot form
inside the inner ILR of the main bar, since the $x_2$ orbits which are
vital for its support do not extend there. In NGC~4303, the secondary
bar is relatively smaller, and a direct calculation of gradients from
measured points of the inner rotation curve gives a position of the
iILR almost coinciding with the inner bar's semi-major axis. If this
is true, the secondary bar in this galaxy is {\it not} made out of
equivalents of the $x_2$ orbits in the outer bar, and gas morphology
and dynamics around it cannot be compared to the model above. However,
the measured inner rotation curve should be treated with caution, as
we are not sensitive to variations in velocity gradient of the inner
4'' due to the resolution of the \CO\ data. As a result, one gets a
rotation curve with almost linear growth for the innermost measured
points. A linear rotation curve generates an angular frequency curve
$\Omega-\kappa/2$ equal identically to zero, and unavoidably creates
an artificial inner ILR. Based on our molecular data, a linear inner
rotation curve cannot be distinguished from a non-linear rotation
curve, which has an $\Omega-\kappa/2$ frequency diverging
monotonically to infinity at small radii, and therefore no inner
ILR. Thus NGC~4303 may not have an iILR at all, which justifies
comparing it to the models of Maciejewski et al. (2002).

\subsection{Scenario (B): A self-gravitating $m=1$ mode? - Formation of a spiral} 

The morphology of the UV sources can also be interpreted as lined up
along a single-arm spiral extending all the way to the galactic
center. In this scenario the effect of the inner bar onto the gas flow
would be negligible. As shown in section \ref{sec:innerkpc}, the
morphology of the star-forming UV clusters in NGC~4303 can be
interpreted as a continuation of the northern \CO\ spiral arm. As there
is no obvious spiral counterpart originating from the southern \CO\ arm,
the UV clusters appear to be arranged into an $m=1$ mode propagating
to the galaxy center. Such lopsidedness is not only detected on large
scales but also in the central parts of spiral galaxies such as
NGC~1808 (Emsellem et al. 2001) and NGC~3504 (Emsellem
2001). Moreover, single-arm nuclear spirals appear to be quite common
in recent high-contrast observations (Martini 2001, Pogge
\& Martini 2002).

No prominent deviations in the residual \CO\ velocity field in the
inner $8''$ of NGC~4303 (Fig.~\ref{fig:resvel}) can be associated with
the UV structure, in contrast to the cases of NGC~1808 and NGC~3504,
where the $m=1$ modes are associated with clear kinematic signatures
(Emsellem 2001). In NGC~4303, the appearance of the southern \CO\
spiral arm is distorted, but this asymmetry in the gas is probably too
weak to display a kinematic signature. If this $m=1$ mode is real, it
underlines the importance of self-gravity for the gas component in the
central region of NGC~4303: The induced and non-self-gravitating $m=1$
mode cannot propagate in nuclear regions (Maciejewski et al. 2002).

Alternatively, the $m=1$ mode may be only a transient feature created
by a single sheared GMC. In Section \ref{sec:innerkpc}, we showed that
one of the more massive GMCs detected in \CO\ might already be able to
form all the UV clusters. Consider a GMC originating from the northern
arm: It gets sheared while spiraling into the galactic center, with
star formation being triggered in separated parts at separate
times. The dynamical time at a radius of 2.5'' ($\sim$ 200~pc) is
10 Myr. The deprojected angle between the northern spiral arm and
the western UV knots is about 120$^o$, which corresponds to a time of
$\sim$ 4 Myr, consistent with the ages derived by Colina \&
Wada (2000). The ages of $\sim$ 10 Myr for the eastern knots are
similar to the travel time of $\sim$ 8 Myr, if one assumes that the
onset of star formation was delayed. As pointed out in
Section~\ref{subsec:kin}, the stellar and molecular gas kinematics
might not be the same, and, therefore, the stellar angular velocity
might be lower, resulting in higher ages. However, in order to test
this hypothesis good knowledge of the stellar kinematics in the inner
6'' as well as age-dating of the individual clusters are essential.

\subsection{Implications for the ''bars-within-bars'' scenario}

In NGC~4303 molecular gas is found (within our spatial resolution) up
to the very nucleus suggesting the presence of a mechanism
transporting the gas down to small radii. However, the nature of this
mechanism is not clear, as the combination of the \CO\ data with the
HST data is consistent with two very different scenarios for the
stellar and the gas component as discussed in the previous
sections. The possible interpretations of our data leave the question
about the importance of the inner bar for the gas flow towards the
center unanswered. If the molecular gas forms a self-gravitating $m=1$
mode or the observed spiral arm is a transient phenomenon, the inner
bar is not important for the fueling of the nucleus. {\it Gaseous}
inner bars as postulated by Shlosman et al. (1989) are characterized
by strong inflowing motions throughout the bar, and the \CO\ emission
being aligned with the bar. Since neither of these are observed, this
scenario has to be rejected. We also find some inconsistency with the
predicted gas flow in stellar double bars: Some UV continuum emission
coincides with the leading side of the inner bar on the scales of the
HST resolution. This suggests that on this smaller scale, the inner
bar in NGC~4303 may enable gas inflow, contrary to the results of the
hydrodynamic model by Maciejewski et al. (2002). Compared to the
modeled bars, the primary bar in NGC~4303 has a stronger potential
which should offset the interaction between the two bars. Also the
ratio of the bar lengths is at the upper end of that observed for
double barred galaxies. This might be expected, since the stronger
primary bar may allow only smaller secondary bars to form within
it. These changes to the potential are likely to alter the gas
flow. Dynamical models exploring a variety of double barred galaxies
as well as more high-resolution observations of double barred galaxies
are essential to better understand the fueling process on scales of a
few 10~pc.

\section{BEYOND DYNAMICAL MODELS}\label{sec:bmods}

\subsection{The asymmetry of the observed gas distribution}\label{sec:gasasym}

The gas leaving the straight shock in hydrodynamic models at the
region where the shock starts to curl enters the diverging 'spray'
flow towards the other shock (e.g. Fig. 8 and 9 of Regan et
al. 1999). This behavior reinforces the bisymmetry of the flow, and
smears out effects of global density variations.

As discussed in Section~\ref{sec:mmobs}, the morphology as well as the
intensity distribution of the \CO\ line emission is not completely
point-symmetric. The western gas lane is about a factor of 1.8
brighter and therefore more massive than the eastern gas lane. Since
we are only detecting about 40\% of the total molecular gas in the
inner arcminute (see Section~\ref{sec:mmobs}), one possible
explanation might be that the asymmetry in intensity is reflecting
global density variations in the now-disturbed molecular gas disk. The
mass/intensity ratio between the northern and southern gas spiral arm
is about 1.6 if the nuclear emission is added to the northern spiral
(assuming this emission is a continuation of the northern spiral, see
Section~\ref{sec:geomdyn}). The clump about $3''$ to the south-east of
the nucleus, between the eastern gas lane and the southern spiral arm,
could then be a result of that asymmetry: dense material from the
western gas lane enters the `spray' region there, and is more visible
than its counterpart from the eastern lane.  The time scale for
molecular gas to travel from the western gas lane to the eastern one
at the radius of $4-8''$ ($310-620$\,pc) can be estimated by assuming
that the rotation velocity is a good measure of the net circular
velocity in this region. We find that it takes only about $5-10$\,Myr.
This would imply that the global density variations were either very
strong in the beginning and the bar potential is still trying to
smooth them out, or that the torque of the bar onto the molecular gas
disk has only started, i.e. the bar has only formed recently. Because
of the large amount of missing flux we favor the second possibility.

\subsection{The role of star formation}\label{subsec:sfmods} 

Most dynamical models do not include star formation and its effects
like supernova explosions and stellar winds. Therefore they might miss
a critical component for the gas dynamics in the inner kiloparsec of
barred galaxies (e.g. Sheth et al. 2000). NGC~4303 offers the
possibility to explore the importance of these effects, as it hosts
young stellar clusters in addition to molecular gas. We neglect the
consumption of molecular gas during star formation, since this effect
is smaller than changes in kinematics resulting from outflows from
young stars.

The molecular gas geometry in the inner $8''$ of NGC~4303 is
disturbed, and shows no tight spatial correlation with the star
formation. This is in contrast to observations of other starburst
rings such as the one in NGC~4314 (Benedict et al. 1996). Although
NGC~4303 is in a tidal interaction with two nearby companions
(NGC~4292 and NGC~4303A, Binggeli et al. 1985), it is very likely that
the gravitational potential of NGC~4303 dominates the inner
kiloparsec. As outlined in section \ref{sec:gasasym}, there is no
obvious reason for the asymmetry from a dynamical point of view. One
possibility might be the impact of massive star formation via stellar
winds and supernovae explosions which will be discussed below.

\subsubsection{The star formation threshold in the ILR region} 

As shown in Section~\ref{sec:innerkpc} the UV continuum lies in the
region of the molecular gas spiral arms, and between the inner and
outer ILR (see section~\ref{sec:ilr}). Elmegreen (1994) used a linear
instability analysis to estimate the critical density for star
formation within the inner Lindblad resonance (ILR). Since it is
possible to measure the gas density in NGC~4303 in the presence of
star formation, we can compare the observations to theoretical
predictions. The critical density $\rho_{crit}$ can be approximated
with $\rho_{crit} = 0.6 \kappa^2 G^{-1}$, where $\kappa$ is the
epicyclic frequency. Using our values for $\kappa (3'') \sim
1050$\,\kms\,kpc$^{-1}$ at the distance of the UV ring of $r \sim
225$\,pc, we get a critical density $\rho_{crit} \approx
140$\,M$_{\odot}$\,pc$^{-3}$. The observed molecular hydrogen mass is
$\sim$ 3.5$\times$10$^7$ \solm. Assuming a ring with a width of about
$1''$ (78\,pc) at a radius of 225\,pc and a gas disk scale height of
30\,pc (see Section~\ref{sec:mmobs}), we find a gas density of
$\sim10$\,M$_{\odot}$\,pc$^{-3}$ for the position of the UV ring. This
is only about a factor of 15 lower than the critical density, similar
to the high critical density found for the starburst ring in M~82
(Wild et al. 1992). However, since the width of the ring is not
resolved in our \CO\ data, the gas density has to be regarded as a
lower limit. Making the ring thinner by a factor of two, similar to
the dust features seen in the HST $V-H$ color map, and correcting the
molecular gas mass for the Helium contribution (36\% of the molecular gas
mass) brings the gas density close enough to the critical density to explain
the presence of the UV starburst ring.

\subsubsection{Feedback of the star formation onto the ISM} 

As pointed out in Section~\ref{sec:innerkpc} the young stellar
clusters might have affected the distribution of the molecular gas by
the impact of stellar winds and supernovae explosions. 
A simple energy estimate can be used to test if such a large
redistribution of the interstellar medium due to stellar winds and SNe
can be responsible for the disturbed \CO\ distribution in the
central kiloparsec of NGC~4303.

The mechanical luminosity (of stellar winds and supernovae) of
a 10$^6$ \solm\ stellar cluster is about 10$^{39.5}$ erg\,s$^{-1}$
during the first few Myr regardless if the star formation is
instantaneous or continuous (e.g. STARBURST99; Leitherer et
al. 1999). Therefore the average stellar cluster in NGC~4303 with a
stellar mass of $0.7 \times 10^5$ \solm\ has released a mechanical
energy of about $3.8 \times 10^{52}$\,erg ($1.1 \times 10^{53}$\,erg)
into the interstellar medium after 6\,Myr (17\,Myr). This is
equivalent to 38 (110) Type~II SNe per stellar cluster.

If the observed molecular gas did coincide with the sites of the
stellar clusters at a radius of $r_{UV}\sim 3''$ (230 pc) at a former
time, then in order to push it to its present location at a radius of
$r_{GMC}\sim 4''$ (310 pc) its potential and kinetic energy have to be
increased. Under the assumption that the gravitational potential can
be approximated by a spherically symmetric mass distribution, the gain
in potential energy is
\begin{equation}
\Delta E_{pot} = m_{GMC} \int_{r_{UV}}^{r_{GMC}} \frac{v_{circ}^2(r)}{r} dr.
\end{equation}
An average cloud of molecular gas in the southern spiral arm has a
mass of $m_{GMC} \sim 5 \times 10^6$ \solm\ (see
Section~\ref{sec:innerkpc}), the velocity at the position of the UV
ring and the southern GMCs are $v(r_{UV}=3'')
\approx 137$ \kms\ and $v(r_{GMC}=4'') \approx 150$ \kms,
respectively. Integration over the circular velocities $v_{circ}$
taken directly from the measured rotation curve (Fig.8) results in the
acquired potential energy of $\Delta E_{pot} \sim 9 \times 10^{53}$\,
erg. At the same time, this cloud of molecular gas will have to gain
the kinetic energy
\begin{equation}
\Delta E_{kin} = m_{GMC} \frac{v(r_{GMC})^2 - v(r_{UV})^2}{2} \approx 2 \times 10^{53} erg
\end{equation}
Thus the total energy needed to move the molecular gas cloud outwards
from $r_{UV}=3''$ to $r_{GMC}=4''$ is the sum of the increment in
potential energy $\Delta E_{kin}$ needed to get out of the potential well,
and the increase in kinetic energy $\Delta E_{kin}$ on the rising part of
the rotation curve. Together these amount to at least $\sim 10^{54}$\, erg.

This value is already considerably larger than the mechanical energy
released by a single stellar cluster during its lifetime. In addition,
most of the mechanical energy heats up the gas, and we did not take
into account these radiative losses. Thus the realistic estimate of
energy needed to push a single molecular cloud out from 3'' to 4'' is
even much higher than $\sim 10^{54}$\, erg; it therefore seems very
unlikely that the UV clusters seen south of the nucleus have altered
the southern CO spiral arm.  

However, a different effect might be important for changing the
apparent gas distribution. Wei\ss\ et al. (2001) have shown in M~82
that the excitation conditions for the molecular lines can differ in
the presence of massive star formation: The molecular gas appears
brighter than the surrounding gas which has lower temperatures, thus
suggesting a different geometry. A multi-transition study of the
molecular gas is needed to test if a similar scenario with varying
excitation conditions can explain the observed asymmetry in
NGC~4303. This highlights the fact that it is essential to better
understand the physical interplay between the molecular gas properties
and massive star formation occurring in the inner kiloparsec of
double-barred galaxies.

\subsection{A forming starburst ring in NGC~4303?}

The striking difference in the molecular gas morphology of NGC~4303
compared to other double barred galaxies such as NGC~4314 (Benedict et
al. 1996) or NGC~1068 (Schinnerer et al. 2000a) is the fact that the
bulk of the molecular gas is located in the gas lanes, and not in the
star-forming ring. This might indicate that the circumnuclear region
in NGC~4303 is in an early state of evolution.  Another difference is
the appearance of the ring: NGC~4314 and NGC~1068 both show almost
circular tightly wound spirals resembling a ring whereas the molecular
gas in NGC~4303 forms a grand design spiral pattern. In addition, the
off-nuclear star formation coincides with the gas rings in NGC~4314
and NGC~1068 suggesting that the stars have formed within the gas
rings. This is in agreement with the assumption that these rings might
be older than the one in NGC~4303. Therefore, the spiral structure
seen in the UV continuum in NGC~4303 might evolve into a ring as well,
similar to those observed in NGC~4314 and NGC~1068. In this case,
on-going star formation is expected to last for a longer time than a
few million years within the ring. This would be in agreement with the
finding of Maoz et al. (2001) who argued for a constant star formation
of about 200\,Myr in the two starburst rings they analyzed.

\section{SUMMARY AND CONCLUSION}\label{sec:conc}

NGC~4303 is one of the first double barred galaxies in which the
molecular gas properties can be studied in detail at high angular and
spectral resolution. The primary bar in NGC~4303 has a typical length
of about 6~kpc whereas the inner bar is considerably smaller, with a
length of about 340~pc. There is a surprisingly good agreement between
the observed overall gas geometry and dynamical models for the gas
flow in barred galaxies.

We detect only about 40\% of the single-dish flux in our
interferometric map, indicating that there is still a large smooth
reservoir of molecular gas in the inner arcminute of NGC~4303. The gas
forms two straight lanes where deviations of up to 90\,\kms\ from the
mean rotational velocity are observed. In the context of hydrodynamic
models this behavior can be explained as inflowing motions in the
large-scale bar. The velocity field of the inner $8''$ is consistent
with a circular rotating disk and shows deviations of only $\sim
5-15$\,\kms. The bulk of the molecular gas is found in the gas lanes
and not in the nuclear region. This may indicate that the
circumnuclear molecular gas is still settling into the typical
ring-like configuration, and that the star forming ring in NGC~4303 is
still young.

The star-forming ring analyzed by Colina et al. (1997, 1999) has no
direct counterpart in the molecular gas distribution. The critical
density for star formation in the UV ring region (region of the ILR)
is only about $\sim15$ times higher than the estimated gas
density. Therefore it seems plausible that the circumnuclear star
formation was started due to the gas inflow triggered by the primary
bar. The disturbed geometry of the southern gas spiral arm suggests
that the massive star formation has affected the molecular gas
distribution. However, a simple energy estimate suggests that the
mechanical energy produced by the stellar clusters is too low to
overcome the gravitational potential. 

There is no strong correlation between the extinction as seen in the HST
$V-H$ map and the distribution of the molecular gas. Possible reasons
are that the light is dominated by different stellar populations as
well as that an ionization cone might have affected the distribution
of extincting material. 

The overall morphology of the molecular gas resembles closely the
predicted gas distribution in dynamical models of strongly barred
galaxies. In the inner kiloparsec the gas lanes start to curl and form
a spiral structure which is distorted in the southern spiral
arm. Detailed comparison between high resolution models of Englmaier
\& Shlosman (2000) and the \CO\ data of NGC~4303 for the inner
kiloparsec shows that the observed spiral structure can be explained
by the density wave generated in a large-scale bar.  No strong
inflowing motions characteristic for a secondary gaseous bar have been
detected in our \CO\ data. After including the HST data in the analysis
of the central 400~pc ($\sim 5''$) two scenarios seem plausible:

\begin{enumerate}

\item The innermost morphology observed with HST shows a nuclear ring 
inside of the nuclear \CO\ spiral: This is uncommon in a single bar, but
consistent with the gas flows in a dynamically possible double bar
model (Maciejewski et al. 2002). The large ratio ($\sim15$) between
the two bar lengths in NGC 4303 is much larger than in this model,
therefore there is room for the \CO\ spiral structure to propagate
inwards before it enters the region dominated by the secondary
bar. The faint UV emission present at the leading side of the inner
bar might indicate inflow of gas along the inner bar.

\item The UV continuum traces an $m=1$ spiral mode which is
self-gravitating, and propagates independently of the inner bar.
\end{enumerate}

It is imperative to study more double barred systems at high angular
resolution to investigate the fueling of the central few
parsecs. Also, dynamical models of double barred galaxies covering a
wide range of different properties such as primary bar strength, and
varying bar length ratios are important in order to better understand
the gas flow in these systems.

This study has shown that the overall gas flow in NGC~4303 is
dominated by the large-scale bar, and very reminiscent of current
dynamical models. However, the observed asymmetry in the gas
distribution in NGC~4303 cannot be explained in detail with current
models. Also, the impact of star formation and feedback on the ISM
inside the ILR may be quite significant, and needs further
consideration. NGC~4303 obviously offers an important example against
which to test elaborate models, though it will be interesting to see
with comparable observations of other double barred galaxies whether
it is unique, or whether it is representative of a broader class of
galaxies.

\acknowledgements

It is a pleasure to thank P. Englmaier, P. Erwin and E. Emsellem for
stimulating discussions. ES acknowledges support by National Sciene
Foundation grant AST 96-13717, and LAM was supported by the PPARC
Rolling Grant PPA/G/O/1999/00193 at the University of Oxford. LAM is
also thankful for partial support by NSF grant AST 96-13717 to visit
the California Institute of Technology as part of this work. Support
of this work was also provided by a grant from the K. T. and
E. L. Norris Foundation. Research with the Owens Valley Radio
Telescope, operated by California Institute of Technology, is
supported by NSF grant AST 96-13717. This research has made use of the
NASA/IPAC Extragalactic Database (NED) which is operated by the Jet
Propulsion Laboratory, California Institute of Technology, under
contract with the National Aeronautics and Space Administration. We
have made use of the LEDA database (http://leda.univ-lyon1.fr).

\clearpage

\begin{figure*}[ht]
\epsscale{.85}
\plotone{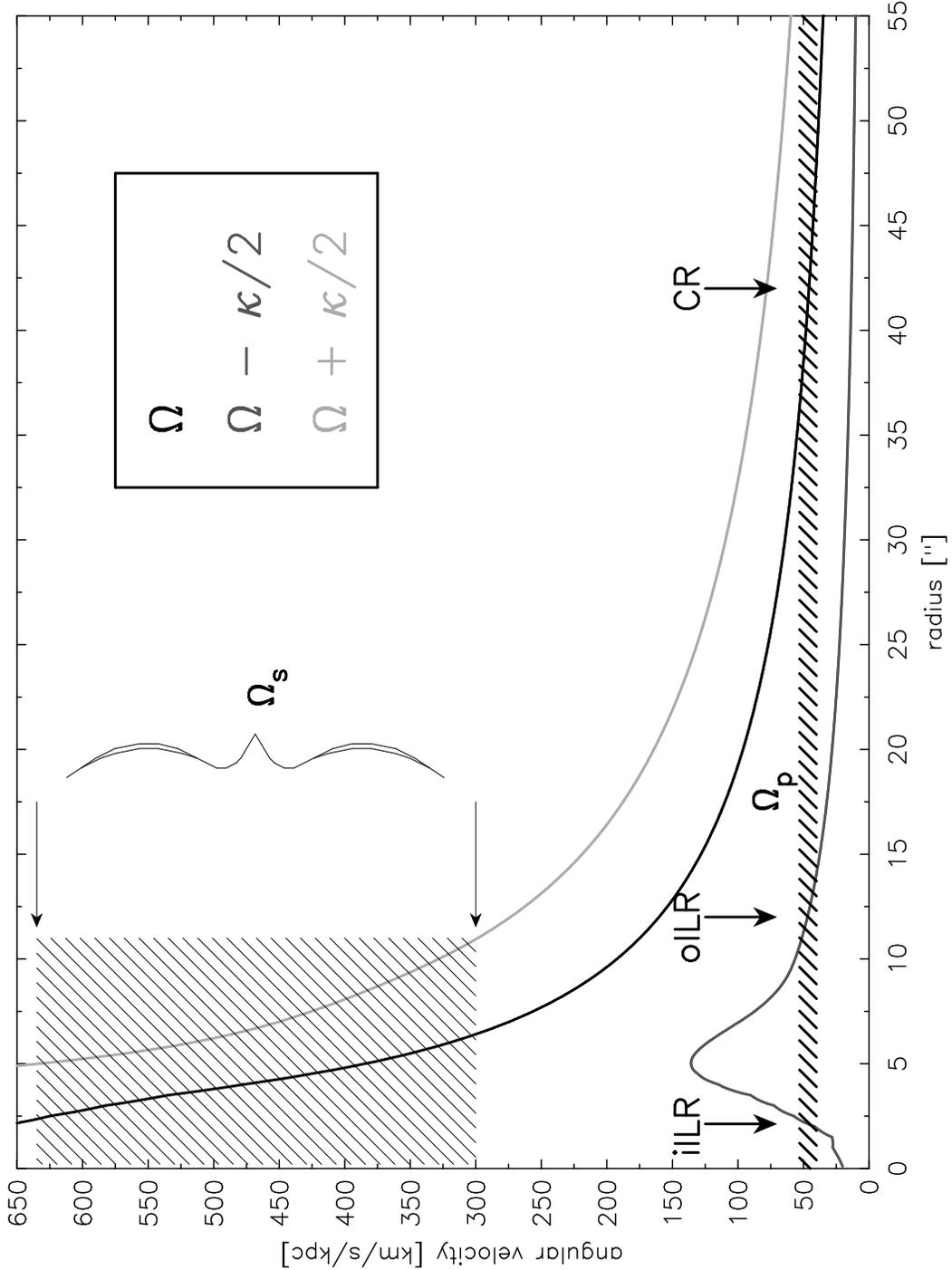}
\caption{
Positions of the dynamical resonances in NGC~4303. Assuming a
corotation resonance (CR) at about $40''$ (see text) implies a
(primary) bar pattern speed of $\Omega_p \sim 40$\kms\,pc$^{-1}$.
Since the rotation velocity is only estimated in the radial range
between $4''$ and $30''$, the position for the outer inner Lindblad
resonance (oILR) of the (primary) bar is roughly at about $12''$. The
position of the inner ILR (iILR), however, is well determined within
the errors. The range for possible pattern speed of the smaller
(secondary) bar $\Omega_s$ is indicated by the hatched area, assuming
that the CR of the secondary bar lies within the ILR of the primary
bar.  }\label{fig:reson}
\end{figure*}

\begin{figure*}[ht]
\epsscale{1.0}
\plotone{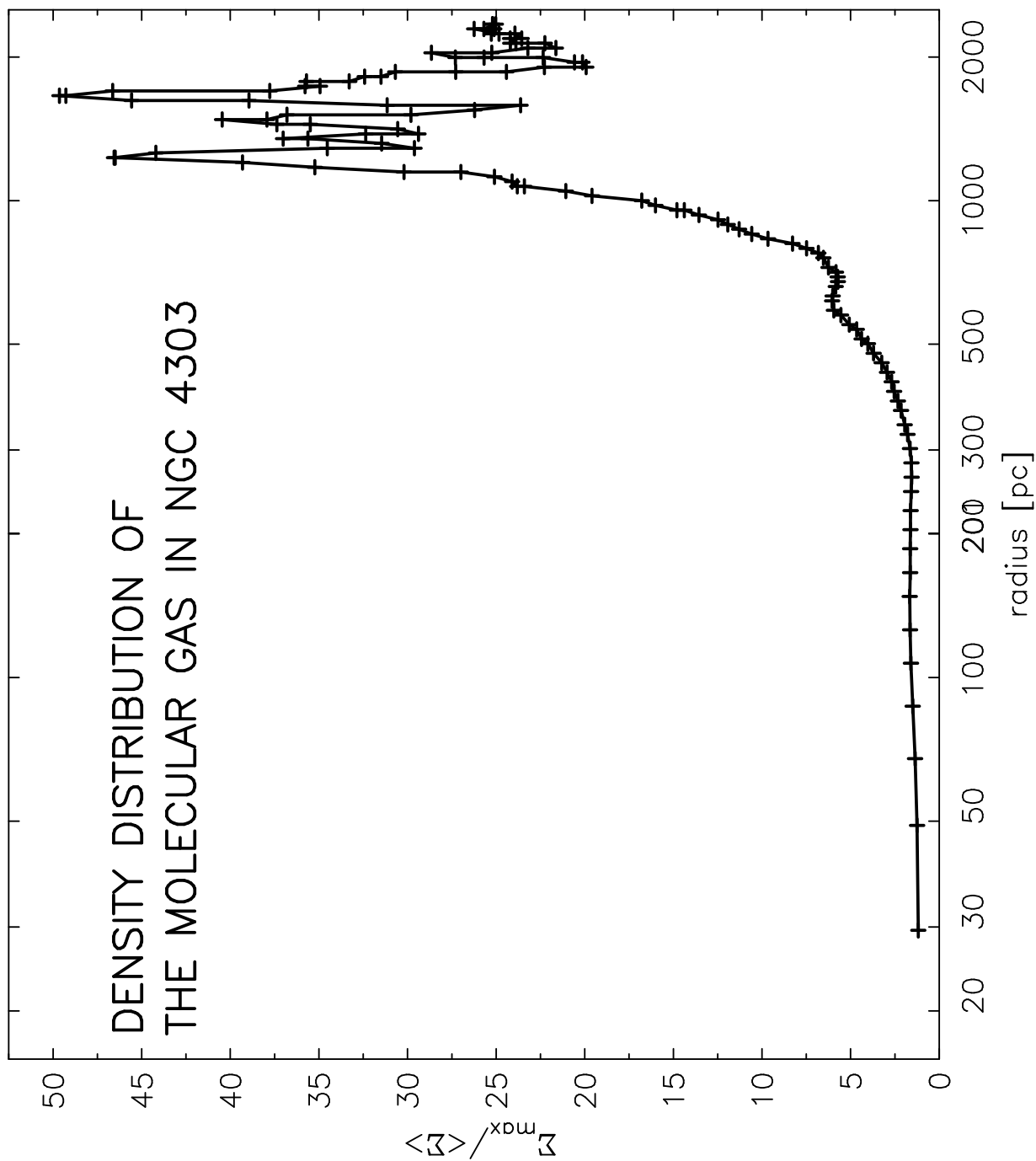}
\caption{
The ratio of the maximum emission and the average emission (derived
from the 0th moment map of the primary beam corrected data) clearly
shows the damping of the shock inside the transition radius $R_t \sim
5''$. This can be compared directly to the density ratio assuming that
the \CO\ emission is proportional to the molecular hydrogen column
density.  }\label{fig:damp}
\end{figure*}

\begin{figure*}[ht]
\epsscale{.60}
\plotone{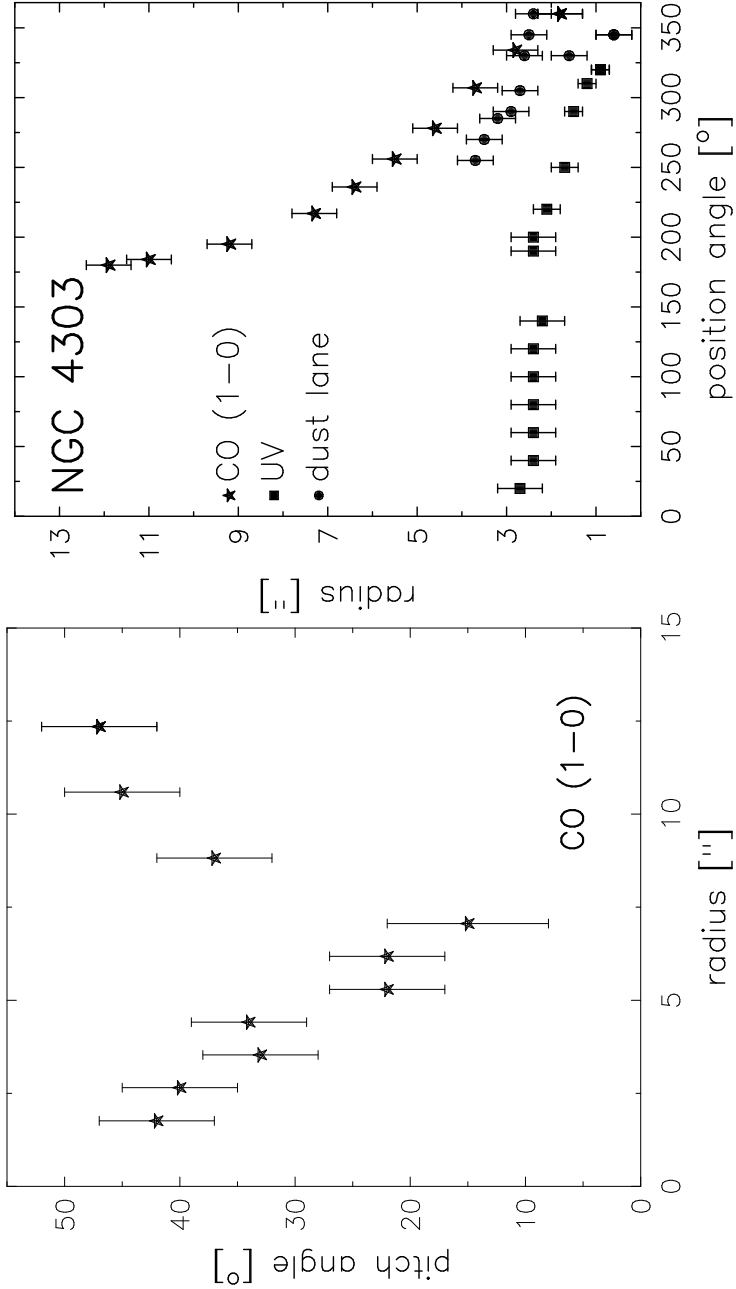}
\caption{
{\it Left:} Pitch angle of the molecular gas spirals (black points) in
the inner $15''$. The angle was measured by eye on the northern spiral
arm in the deprojected deconvolved intensity map (see
Fig. \ref{fig:logsp}). {\it Right:} A change of radius with position
angle is also indicating a spiral pattern. The radius was measured for
different position angle in the deprojected images for the northern CO
spiral arm (stars), the UV clusters (squares), and the dust lanes seen
in the $V-H$ map (circles). The position angle starts at the
(deprojected) major axis and is counted clock-wise. The UV clusters at
a PA of $0^o$ form a continuation of the dust lane at PA of $360^o$.
For the position angles of $330^o$ and $345^o$, we give two measured
points for the dust lanes, since the dust lane splits into two parts
for these angles. See Section \ref{subsec:gashst} for further discussion.
}\label{fig:pitch}
\end{figure*}

\begin{figure*}[ht]
\plotone{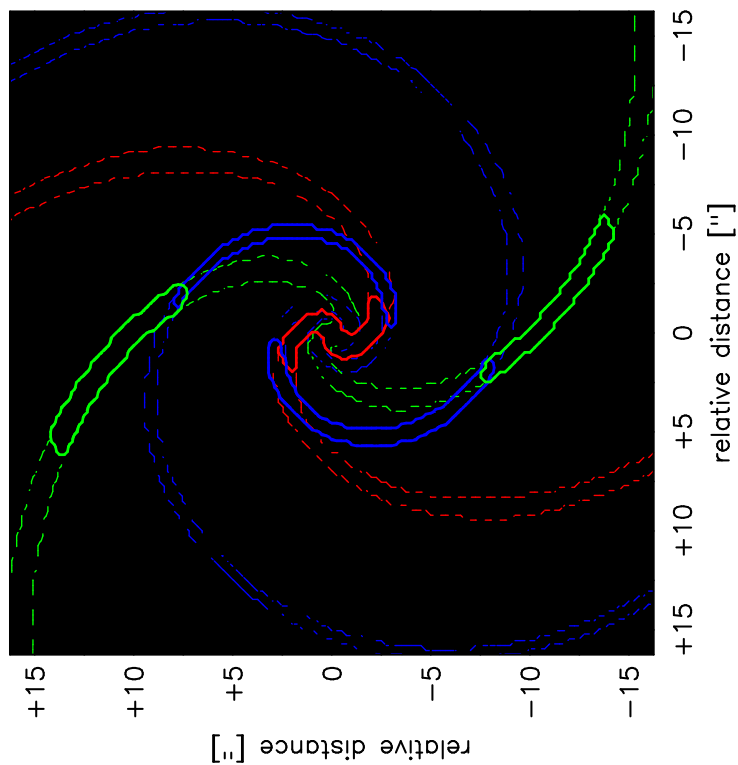}
\caption{ 
Logarithmic spirals with fixed pitch angles of $40^{\circ}$ (red
contours), $20^{\circ}$ (blue contours) and $45^{\circ}$ (green
contours) are overlaid on the deprojected \CO\ 0th moment map
(gray-scale). The part of the logarithmic spirals which was fitted to
the \CO\ data is in solid lines whereas the rest is shown in broken
lines. The deconvolved (LUCYed) map which was used to fit the pitch
angle is shown in black contours. The required change of the pitch
angle for various radii (see Fig. \ref{fig:pitch}) is quite
obvious. The major kinematic axis is now running north-south due to
the deprojection. This implies that the northern spiral arm is now the
eastern one.  }\label{fig:logsp}
\end{figure*}

\clearpage

\begin{table}[htb]
\caption{
Summary of the HST Archival data of the nuclear region of
NGC~4303.}\label{tab:hstlog} 
\begin{center}
\begin{tabular}{llcr}\hline \hline
Dataset & Filter & $\lambda_{cent}$ [$\mu m$] & Exposure [$s$] \\
\hline \hline
   &  NUV-MAMA  & 0.2  &  $3\times520$ \\
   &  F606W  & 0.6  &  $2\times160$ \\
   &  F160W  & 1.6  &  $320$          \\
\hline \hline
\end{tabular}
\end{center}
\end{table}

\begin{table}[htb]
\caption{
\label{ww1}
Some properties of NGC 4303 (= M~61)}
\begin{center}
\begin{tabular}{lc}\hline \hline
 & NGC~4303  \\ \hline 
Right ascension (J2000) & 12$^h$ 21$^m$ 54.99$^s$ \\
Declination (J2000) & 04$^{\circ}$ $28'$ $25.55''$ \\
Classification & SAB(rs)bc \\
Inclination & 25$^{\circ}$ \\
Position angle & 318$^{\circ}$ \\
AGN type & LINER/Sey~2 \\
Systemic velocity $v_{\rm LSR}$& 1560\,\kms\\
Distance & 16.1 Mpc \\
$1''$ equals & 78~pc \\
\hline \hline
\end{tabular}
\end{center}
The sky coordinates, the systemic velocity, and the AGN type
were taken from NED (NASA/IPAC Extragalactic Database). The
classification is from the RC3 catalog (de Vaucouleurs et al. 1991).
Inclination and position angle are described in
Section~\ref{subsec:dyn}.  For the distance we adopted the value of
M~100, the largest spiral in the Virgo cluster ($16.1$\,Mpc, Ferrarrese
et al. 1996).
\end{table}

\begin{table}[htb]
\caption{
$^{12}$CO\ fluxes and Molecular Gas Masses}\label{tab:COfluxes}
\begin{center}
\begin{tabular}{lrrrr}\hline \hline
Component &    I$_{CO}$  &  S$_{CO}$   &  N$_{H_2}$   &  M$_{H_2
}$   \\
          & [Jy \kms]& [K \kms]& [10$^{22}$cm$^{-2}$]& [10$^7$ M$_{\odot}$]\\ 
\hline
Northern peak             &  18.1   &  420   &  8.4 & 2.4  \\
Southern peak             &   5.7   &  131   &  2.6 & 0.8  \\
western gas lane (north)  &   3.4   &   80   &  1.6 & 0.5  \\
western gas lane (south)  &  53.3   & 1240   & 24.7 & 7.2  \\
eastern gas lane (north)  &  30.1   &  700   & 13.9 & 4.0  \\
eastern gas lane (south)  &   2.4   &   55   &  1.1 & 0.3  \\
Northern spiral arm       &  26.5   &  615   & 12.3 & 3.6  \\
Southern spiral arm       &  14.3   &  330   &  6.6 & 1.9  \\
Nucleus                   &   6.9   &  160   &  3.2 & 0.9  \\
Nuclear Disk              &  51.2   & 1190   & 23.8 & 6.9  \\
\hline \hline
\end{tabular}
\end{center}
The \COe\ line fluxes and molecular gas masses for various components
of the molecular gas distribution. The components are indicated in
Fig.~\ref{fig:COfet}. Note, that the molecular gas masses have an
uncertainty of a factor of about 2 - 3 due to the uncertainty in the
\nhico\,conversion factor.
\end{table}


\clearpage

\begin{thebibliography}

\bibitem{} Athanassoula, E., 1992, \mnras, 259, 345; (A92)
\bibitem{} Banfi, M., Rampazzo, R., Chincarini, G., Henry, R.B.C., 1993, \aap, 280, 373
\bibitem{} Benedict, G.F., Smith, B.J., Kenney, J.D.P., 1996, \aj, 112, 1318
\bibitem{} Binggeli, B., Sandage, A., Tammann, G.A., 1985, \aj, 90, 1681
\bibitem{} Buta, R., Crocker, D.A., 1993 \aj, 105, 1344
\bibitem{} Calzetti, D., Kinney, A.~L., \& Storchi-Bergmann, T.\ 1994, \apj, 429, 582 
\bibitem{} Cayatte, V., van Gorkom, J.H., Balkowski, C., Kotanyi, C., 1990, \aj, 100, 604
\bibitem{} Chapelon, S., Contini, T., Davoust, E., 1999, \aap, 345, 81
\bibitem{} Contopoulos, G., 1981, \aap, 102, 265
\bibitem{} Colina, L., Arribas, S., 1999, \apj, 514, 637
\bibitem{} Colina, L., Garcia Vargas, M.L., Mas-Hesse, J.M., Alberdi, A., Krabbe, A., 1997, \apj, 484, L41
\bibitem{} Colina, L., Wada, K., 2000, \apj, 529, 845
\bibitem{} Combes, F., Becquaert, J-F., 1997, \aap, 326, 554
\bibitem{} Elmegreen, B.G., 1994, \apjl, 425, L73
\bibitem{} Elmegreen, B.G., Efremov, Y., Pudritz, R.E., Zinnecker, H. 1999, in ''Protostars and Planets IV'', Univ. of Arizona Press, p. 179 
\bibitem{} Emsellem, E., 2001, in ''The Central kpc of Starbursts and AGN'', ed. J.H. Knapen, J.E. Beckman, I. Shlosman, \& T.J. Mahoney, ASP Conf. Ser., 249, 91
\bibitem{} Emsellem, E., Greusard, D., Combes, F., Friedli, D., Leon, S., P{\'e}contal, E., Wozniak, H.\ 2001, \aap, 368, 52 
\bibitem{} Englmaier, P., Gerhard, O., 1997, \mnras, 287, 57
\bibitem{} Englmaier, P., Shlosman, I., 2000, \apj, 528, 677
\bibitem{} Erwin, P., Sparke, L.S. 2002, \aj, in press, astro-ph/0203514
\bibitem{} Ferrarese, L., et al. 1996, ApJ, 464, 568
\bibitem{} Frei, Z., Guhathakurta, P., Gunn, J.E., Tyson, J.A., 1996, \aj, 111, 174
\bibitem{} Friedli, D. 1996, \aap, 312, 761
\bibitem{} Friedli, D., Martinet, L., 1993, \aap, 277, 27
\bibitem{} Garcia, A.M., 1993, \aap Suppl., 100, 47
\bibitem{} Genzel, R., Weitzel, L., Tacconi-Garman, Blietz, M., Krabbe, A., Lutz, D., Sternberg, A., 1995, \apj, 444, 129
\bibitem{} Guhathakurta, P., van Gorkom, J.H., Kotanyi, C.G., Balkowski, C., 1988, \aj, 96, 851
\bibitem{} Heraudeau, Ph., Simien, F., 1998, \aap Suppl., 133, 317
\bibitem{} Ho, L.C., Filippenko, A.V., Sargent, W.L.W., 1997, \apj, 487, 591
\bibitem{} van der Hulst, J.M., Terlouw, J.P., Begeman, K., Zwitser, W., 
   Roelfsema, P.R., in ``Astronomical Data Analysis Software and Systems I'', 
   eds. D. M. Worall, C. Biemesderfer and J. Barnes, 
   ASP Conf. series no. 25, p. 131
\bibitem{} Jogee, S., Shlosman, I., Laine, S., Englmaier, P., Knapen, J.H., Scoville, N.Z., Wilson, C.D., 2002, \apj, sub., astro-ph/0202270
\bibitem{} Jungwiert, B., Combes, F., Axon, D.J. 1997, \aap Suppl., 125, 479
\bibitem{} Kenney, J.D., Young, J.S., 1988, \apjs, 66, 261
\bibitem{} Koopmann, R.A., Kenney, J.D.P., Young, J., 2001, \apjs, 135, 125
\bibitem{} Laine, S., Knapen, J.H., P{\' e}rez-Ram{\' i}rez, D., Englmaier, P., Matthias, M.\ 2001, \mnras, 324, 891
\bibitem{} Laine, S., Shlosman, I., Knapen, J.H., Peletier, R.F. 2002, \apj, 567, 97
\bibitem{} Leitherer, C. et al. 1999, \apjs, 123, 3 
\bibitem{} Ma, J., Peng, Q.-H., Gu, Q.-S., 1998, \aap Suppl., 130, 449
\bibitem{} Maciejewski, W., Sparke, L.S., 1997, \apj Letters, 484, L117
\bibitem{} Maciejewski, W., Sparke, L.S., 2000, \mnras, 313, 745
\bibitem{} Maciejewski, W., Teuben, P.J., Sparke, L.S., Stone, J.M. 2002, \mnras, 329, 502
\bibitem{} Maoz, D., Barth, A.J., Ho, L.C., Sternberg, A., \& Filippenko, A.V.\ 2001, \aj, 121, 3048
\bibitem{} Martin, P., 1995, \aj, 109, 2428
\bibitem{} Martini, P., 2001, in ''The Central kpc of Starbursts and AGN'', ed. J.H. Knapen, J.E. Beckman, I. Shlosman, \& T.J. Mahoney, ASP Conf. Ser., 249, 98
\bibitem{} Martini, P., Pogge, R.W. 1999, \aj, 118, 2646
\bibitem{} Meier, D.S., Turner, J.L., Hurt, R.L. 2000, \apj, 531, 200
\bibitem{} Perez-Ramirez, D., Knapen, J.H., Peletier, R.F., Laine, S., Doyon, R.,
    Nadeau, D., 2000, \mnras, 317, 234
\bibitem{} Piner, B.G., Stone, J.M., Teuben, P.J., 1995, \apj, 449, 508
\bibitem{} Pogge, R.W., Martini, P., 2002, \apj, in press, astro-ph/0201185
\bibitem{} Quillen, A.C., De Zeeuw, P.T., Phinney, E.S., Phillips, T.G., 1992, \apj, 391, 121
\bibitem{} Rautiainen, P., Salo, H.\ 1999, \aap, 348, 737 
\bibitem{} Regan, M.W., Mulchaey, J.S., 1999, \aj, 117, 2676
\bibitem{} Regan, M.W., Sheth, K., Vogel, S.N., 1999, \apj, 526, 97
\bibitem{} Rubin, V.C., Waterman, A.H., Kenney, J.D.P., 1999, \aj, 118, 236
\bibitem{} Sault, R.J., Teuben, P.J., Wright, M.C.H., 1995, in ``Astronomical Data
    Analysis Software and Systems IV'', ed. R. Shaw, H.E. Payne, J.J.E. Hayes,
    ASP Conf. Ser., 77, 433
\bibitem{} Schinnerer, E., Eckart, A., Tacconi, L.J., 2000b, \apj, 533, 850
\bibitem{} Schinnerer, E., Eckart, A., Tacconi, L.J., Genzel, R., Downes, D. 2000a, \apj, 533, 850
\bibitem{} Schmitt, H.R., Kinney, A.I., 1996, \apj, 463, 498
\bibitem{} Scoville, N.Z., Carlstrom, J.E., Chandler, C.J., Phillips, J.A., Scott, 
    S.L., Tilanus, R.P.J., Wang, Z., 1993, \pasp, 105, 1482
\bibitem{} Sheth, K., 2001, Ph.D. thesis, Univ. of Maryland
\bibitem{} Sheth, K., Regan, M.~W., Vogel, S.~N., \& Teuben, P.~J.\ 2000, \apj, 532, 221
\bibitem{} Shlosman, I., Frank, J., Begelman, M.C. 1989, Natur, 338, 45
\bibitem{} Strong, A. W., et al. 1987, Proc. 20th Intern. Cosmic Ray Conf., I, 125
\bibitem{} Telesco. C.M., Becklin, E., Wynn-Williams, C., Harper. D., 1984, \apj, 282, 427
\bibitem{} Wada, K., Norman, C.A., 1999, \apjl, 516, L13
\bibitem{} Warmels, R.H., 1988, \aap Suppl., 72, 57
\bibitem{} Wei\ss, A., Neininger, H\"uttemeister, S., Klein, U., 2001, \aap, 365, 571
\bibitem{} Wei\ss, A., Walter, F., Neininger, N., Klein, U., 1999, \aap, 345, L23
\bibitem{} Wild, W., Harris, A.I., Eckart, A., Genzel, R., Graf, U.U., Jackson, J.M., Russell, A.P.G., Stutzki, J.\ 1992, \aap, 265, 447 
\bibitem{} Young, J.S., et al., 1995, \apjs, 98, 219

\end{thebibliography}
\end{document}